\documentclass[a4paper,11pt]{article}
\pdfoutput=1 
\usepackage{jcappub}
\usepackage{mathtools}
\usepackage{graphicx}
\usepackage{natbib}
\usepackage{multirow}
\usepackage{amsmath}
\usepackage[amssymb]{SIunits}
\usepackage{empheq}
\usepackage{booktabs}

\usepackage{color}

\newcommand{\TM}{T_{\textrm{M}}}
\newcommand{\TCMB}{T_{\textrm{CMB}}}

\usepackage[T1]{fontenc} 

\title{\boldmath Cosmological constraints on exotic injection of electromagnetic energy}

\author[a,b]{Vivian Poulin,}
\author[b]{Julien Lesgourgues,}
\author[a,b]{Pasquale D. Serpico}

\affiliation[a]{LAPTh, Universit\'e Savoie Mont Blanc \& CNRS,\\ 9 Chemin de Bellevue BP 110 Annecy-le-Vieux F-74941 Annecy Cedex, France.}
\affiliation[b]{Institute for Theoretical Particle Physics and Cosmology (TTK), \\ RWTH Aachen University, D-52056 Aachen, Germany.}

\emailAdd{Vivian.Poulin@lapth.cnrs.fr}
\emailAdd{Pasquale.Serpico@lapth.cnrs.fr}
\emailAdd{Julien.Lesgourgues@physik.rwth-aachen.de}

\abstract{We compute cosmic microwave background (CMB) anisotropy constraints on exotic forms of energy injection in electromagnetic (e.m.) channels over a large 
range of timescales.  We show that these constraints are very powerful around or just after recombination,
although CMB keeps some sensitivity e.g. to decaying species with lifetimes as long as $10^{25}\,$s. These bounds are complementary to CMB spectral distortions and primordial nucleosynthesis ones, which dominate at earlier timescales, as we also review here.   
For the first time, we describe the effects of the e.m. energy injection on the CMB power spectra {\it as a function of the injection epoch}, using the lifetime of a decaying particle as proxy. We also identify a suitable {\it on-the-spot} approximation, that can be used to derive accurate constraints, and describe its differences with the most up-to-date treatment. Our results are of interest not only for early universe relics constituting (a fraction of) the dark matter, but also for other exotic injection of e.m. radiation. For illustration, we apply our formalism to: i) Primordial black holes of mass $10^{13.5}\,{\rm g}\lesssim M\lesssim 10^{16.8}\,$g,  
 showing that the constraints are comparable to the ones obtained from gamma-ray background studies and even dominate below $\sim10^{14}\,$g. ii) To a peculiar mass-mixing range in the  sterile neutrino parameter space, complementary to other astrophysical and laboratory probes. 
iii) Finally, we provide a first estimate of the room for improvement left for forthcoming 21 cm experiments, comparing it with the reach of proposed  CMB spectral distortion ({\sf PiXiE}) and CMB angular power spectrum ({\sf CORE}) missions. We show that the best and most realistic opportunity to look for this signal (or to improve over current constraints) in the 21 cm probe is to focus on the Cosmic Dawn epoch, $15\lesssim z\lesssim30$, where the qualitatively unambiguous signature of a spectrum {\it in emission} can be expected for models that evade all current constraints.}

\begin{document}
\hfill{\small LAPTH-063/16, TTK-16-45}\\
\maketitle
\flushbottom
\section{Introduction}
Besides helping us reconstructing the evolution of the universe, cosmological probes also provide us with a historical `archive' of the particles populating the early phases of the
cosmos, as well as informing us about relevant interactions that took place. For instance, primordial (or `Big Bang') nucleosynthesis (BBN) or the cosmic microwave background (CMB), are sensitive to the standard model neutrinos populated via weak interactions, as well as to the amount of baryons present in the primordial plasma, generated by yet unknown processes beyond the standard model. 
We have also cosmological evidence for a sizable dark matter (DM) component, whose nature is however still unknown. It is even possible for instance that light and ``dark'' relativistic particles are present in the cosmic soup, whose cosmological properties are partially degenerate with neutrino ones,  or that DM is made of different components with wildly different nature and properties such as coupling with ordinary matter or lifetime.

Numerous relics from the early universe have been proposed in many extensions of the standard model of particle physics, in some cases unstable to processes injecting electromagnetic (e.m.) forms of energy. For instance, decaying massive particles can be the progenitors of the DM in `superWIMP' models~\cite{Feng:2004zu}, or the DM itself can be unstable, provided its lifetime is much longer than the lifetime of the universe, as in R-parity breaking SUSY models (see e.g.~\cite{Bouquet:1986mq}). Right-handed massive neutrinos commonly invoked in  scenarios to give masses to (active) neutrinos are also unstable, with very long lifetimes if light enough and/or with tiny mixing angles.
Primordial black holes (PBHs), if sufficiently light, also evaporate via Hawking radiation over timescales not too long compared with the age of the universe (for a review, see \cite{Carr:2009jm}). Our considerations also apply to an excited DM component~\cite{Finkbeiner:2007kk}, injecting energy when the metastable state de-excites. All these models typically lead to photons and leptons among their final states, with peculiar time dependences due to the model considered. As we will review in the following, BBN and/or CMB spectral distortions provide constraints to the amount of energy injected at early epochs, while CMB anisotropy considerations become more relevant for longer lifetimes. 
It is important to assess the relative constraining power of these different tools, which is the main task we address in the following.
Also, forthcoming 21 cm surveys promise to open a window on the yet unconstrained ``Dark Ages''. It is interesting to estimate whether this probe has the potential to improve over current CMB constraints. For DM annihilation, it has been shown recently \cite{Lopez-Honorez:2016sur} that, as long as astrophysical uncertainties on star formation dominate, CMB anisotropies would have a better sensitivity than 21 cm. However, the situation may be different for exotic physics entering the game at specific ages, as we will discuss in the following.

In section \ref{sec:CMBconstraints}, we focus on CMB power spectra constraints, describing for the first time the effects of the e.m. energy injection onto the CMB power spectra {\it as a function of the lifetime} of the decaying particles. In section \ref{sec:OtherConstraints}, we review how to compute spectral distortions and BBN constraints, which typically are the dominante ones when the lifetime is below $10^{12}\,$s, whereas CMB anisotropies are very powerful at constraining energy injection after this time. 
The lifetime dependence of the CMB anisotropy bounds that we derived also inspired us in proposing and testing an {\it on-the-spot} approximation to the more accurate treatment, whose accuracy we tested. In section \ref{sec:Applications}, we apply our treatment to a few candidates of interest: first, we deal with PBHs of relatively low mass, $10^{13.5}\,{\rm g}\lesssim M\lesssim 10^{16.8}\,$g, that can affect CMB power spectra due to their evaporation. To our knowledge, realistic constraints on this scenario have never been computed so far, yet our results show that the bounds are competitive with existing ones. Then, we show the application of our results to sterile neutrinos with keV-MeV mass scale. Finally, 
we provide a first estimate of the room for improvement left for forthcoming 21 cm experiment SKA~\footnote{https://www.skatelescope.org/}, comparing it with the reach of a proposed   satellite targeting CMB spectral distortions ({\sf PiXiE}~\cite{Kogut:2011xw}) and the CMB angular power spectrum accuracy achievable with a next-generation satellite ({\sf CORE}-like~\cite{Bouchet:2011ck}). We show that the best and most realistic opportunity to look for this signal (or to improve over current constraints) in the 21 cm probe is to focus on the Cosmic Dawn epoch, $15\lesssim z\lesssim30$, where the qualitatively unambiguous signature of a spectrum {\it in emission} can be present, with negligible effects expected from astrophysical sources. We finally report our conclusions in section \ref{sec:Concl}.
\section{CMB power spectra constraints}\label{sec:CMBconstraints}

In this section, we describe our treatment of the injection of electromagnetic energy in the cosmological medium and its effects on the CMB power-spectra.
Although the injection history is model dependent, a useful proxy is to assume the decay of an
exotic `particle'. Since this also describes the scenario most commonly found in the literature, henceforth we will thus refer to a decaying particle.
Our considerations are however more generic than that, as we will show later on with a specific example.  
Also, we neglect gravitational effects, i.e. we assume that only a small fraction of the total dark matter density is decaying (or that its lifetime 
is much longer than the age of the universe, if its abundance is not small). In general, whenever an e.m. decay channel is non-negligible, the bounds obtained are many orders of magnitude 
stronger than the purely gravitational ones  (see refs~\cite{Audren:2014bca,Poulin:2016nat} for recent treatments). The main effect of the decay byproducts is then their impact on the free electron fraction, which in turn affects CMB anisotropy angular spectra. The separation of scales between purely gravitational effects and e.m. ones also allow us to isolate the latter, described in details in what follows.

\subsection{Standard equations\label{sec:standard_eq}}
 Both {\it stable} and {\it unstable} standard model (anti)particles are found among the final states of the exotic particle decay. The latter  decay into standard model stable (anti)particles very quickly compared to other relevant timescales, hence one can limit oneself to treat only the energy deposition problem of stable standard model particles and antiparticles. These particle are either ``inert'', simply losing energy adiabatically via redshift, or interact with the gas (primordial plasma) after (before) recombination, transferring their energy to the photons and light element atoms (electrons and nuclei).  It is standard to neglect the energy deposited by protons/antiprotons and neutrinos: The latter are basically invisible to the medium and simply carry away part of the energy, hence neglecting them is an excellent approximation. Ignoring the energy deposition by protons and antiprotons has also been checked to loosen the bounds at the 10\% level \cite{Weniger:2013hja}. Neglecting those processes thus leads only to modestly conservative bounds, while permitting a significant reduction of the computing time. As a result, the deposition process essentially concerns energetic photons and electrons/positrons~\cite{Chen}. Their injection in the medium initiates the development of a high energy electromagnetic (e.m.) cascade, which proceeds in the following schematic way: i) the number of  non-thermal particles grows, while their average energy decreases, mostly due to interactions with relic photons; ii) when the non-thermal particles energy reach the keV range, they start interacting with atoms, mostly hydrogen. Some interaction also involves helium atoms (but this has been shown to be a sub-leading, often negligible effect~\cite{Galli13}) and free electrons. An accurate description of the evolution of the daughter particle spectra over many energy and time scales is necessary in order to correctly capture the energy deposition process. Indeed, it has been shown that at redshift around and below recombination the injected energy is often not absorbed {\em on-the-spot}, rather can redshift away before being deposited~\cite{Slatyer09}. Dedicated numerical tools have been developed to deal with the relevant physical processes, 
and we shall make use of the results from Ref.~\cite{Slatyer09} recently updated in Ref.~\cite{Slatyer15-1}.
The main e.m. impact of exotic particle decay is to modify  the fraction of free electrons $x_e$, either through direct ionization or collisional excitation followed by photoionization by a CMB photon. An indirect effect is via the heating of the intergalactic medium (IGM), whose temperature we denote with $\TM$, and which has a feedback on the evolution of $x_e$. 
In turn, through interactions of the CMB photons with free electrons, these processes will have an impact on the CMB anisotropy angular power spectra. 
In order to follow the evolution equations for $x_e$ and $\TM$ we use the numerical code {\sf Recfast} \cite{Seager:1999bc} v1.5 as implemented in the Boltzmann code {\sf CLASS}\footnote{\tt http://class-code.net} \cite{Lesgourgues:2011re,Blas:2011rf} v2.5.
In this code, the evolution of the free electron fraction is ruled by a system of coupled differential equations of the type\footnote{In reality {\sf Recfast} contains equations that are modified with fudge factors calibrated on more accurate code such as {\sf CosmoRec} \cite{Chluba:2005uz,Chluba:2010ca} and {\sf HyRec} \cite{AliHaimoud:2010dx}. It is not necessary to go beyond the use of {\sf Recfast} as long as the ionization history around recombination, for which high precision calculation is mandatory, is not too far away from $\Lambda$CDM. Since large energy injections around recombination are ruled out, it is safe for us to work with {\sf Recfast} only. We have checked this explicitly by comparing results with the last public version of {\sf HyRec}.}
\begin{eqnarray}\label{eq:x_e&T_M}
\frac{dx_{e}(z)}{dz}=\frac{1}{(1+z)H(z)}(R(z)-I(z)-I_X(z))~,\nonumber\\
\frac{d\TM}{dz}  =  \frac{1}{1+z}\bigg[2\TM+\gamma(\TM-\TCMB)\bigg]+K_h~.
\end{eqnarray}
where the $R$ and $I$ terms are the standard recombination and ionization rates given by
\begin{equation}\label{eq:RandI}
R(z) = C\bigg[\alpha_{H} x_e^2 n_H\bigg], \qquad I(z)  = C\bigg[ \beta_{H}(1-x_e)e^{-\frac{h\nu_\alpha}{k_b\TM}}\bigg].
\end{equation}
 $I_X(z) = I_{Xi}(z)+I_{X\alpha}(z)$ is an effective ionization rate where the rate of direct ionization $I_{Xi}$ and excitation+ionization $I_{X{\alpha}}$ are given by:
\begin{equation}\label{eq:IonizationRate}
I_{Xi} = -\frac{1}{n_H(z)E_i}\frac{dE}{dVdt}\bigg |_{\textrm{dep}, i}~, \quad I_{X{\alpha}} =  -\frac{(1-C)}{n_H(z)E_{\alpha}}\frac{dE}{dVdt}\bigg |_{\textrm{dep}, \alpha}~,
\end{equation}
where $E_i$ and $E_\alpha$ are respectively the average ionization energy per baryon, and the Lyman-$\alpha$ energy.
Finally, the rate $K_h$ at which DM decays or annihilations heat the plasma is defined as:
\begin{equation}\label{eq:Kh}
K_h =-\frac{2}{H(z)(1+z)3k_b n_H(z)(1+f_{He}+x_e)}\frac{dE}{dVdt}\bigg |_{\textrm{dep}, h}\quad.
\end{equation}
We refer to appendix of Ref.~\cite{Poulin:2015pna} for further definitions and more details on each coefficient.\\
The energy \emph{deposited} in the plasma at redshift $z$, $\frac{dE}{dVdt} \big |_{\textrm{dep}}$ is splitted between ionization, excitation of the  Lyman-$\alpha$ transition, heating and very low energy photons ($\lesssim$10.2 eV) unable to interact.
In case of a decaying particle with lifetime $\tau$, the rate of energy injection per unit volume is given by 
\begin{equation}\label{eq:EnergyInjected}
\frac{dE}{dVdt}\bigg |_{\textrm{inj}}=(1+z)^3\Xi\,\Omega_{\rm DM}\rho_c c^2\Gamma\,e^{-\Gamma\,t}\quad,
\end{equation}
where  $\rho_c$  is the current critical density, $\Gamma$ is the width (inverse lifetime), $\Xi$ is the relative amount of energy released into e.m. for a single decay, arbitrarily normalized to the current total cold DM abundance, $\Omega_{\rm DM}$. For instance, a species constituting 1\% of the total DM abundance decaying into $\nu\gamma$ corresponds to $\Xi=1/200$.
We follow the by now standard method of Refs.~\cite{Finkbeiner11,Slatyer12} to take into account e.m. energy injection in the periods concerned.  In that case, the deposited energy is related to the injected one by:
\begin{equation}
\frac{dE}{dVdt}\bigg |_{\textrm{dep},c}=f_c(z,x_e)\frac{dE}{dVdt}\bigg |_{\textrm{inj, long-lived}}\quad.
\end{equation}
where the subscript $c$ denotes the ``channel'' (ionization, excitation,\ldots). Here, ${\frac{dE}{dVdt}}|_{\textrm{inj, long-lived}}$ corresponds to the injection rate in case of a long-lived particle and the exponential factor $e^{-\Gamma\,t}$ is absorbed in the definition of the $f_c(z,x_e)$ functions. These functions encode all the physics of the energy deposition and we will describe them in the following. In principle, transport equations accounting for all standard model electromagnetic processes (from high-energy QED ones to atomic processes) allow one to compute the functions  $f_c(z,x_e)$  ``case by case''. In practice, at very least for computing time limitations, some simplifications are needed if one wants to study a large range of models.
For instance, it is standard to perform the factorization approximation 
\begin{equation}\label{eq:factorization}
f_c(z,x_e)\simeq f(z)\,\chi_c(x_e)
\end{equation}
which has been shown to work very well for  injection energies $\gtrsim$10 MeV \cite{Slatyer15-2} and that corresponds to a factorization between high-energy processes (determining $f(z)$) and low-energy processes, approximately universal, responsible for the absorption repartition fractions $\chi_c(x_e)$.
The $\chi_c(x_e)$ functions have been computed in several references, the most recent one being Ref.~\cite{Galli13}, which we adopt. 

\begin{figure}[t]
\centering
\includegraphics[scale=0.3]{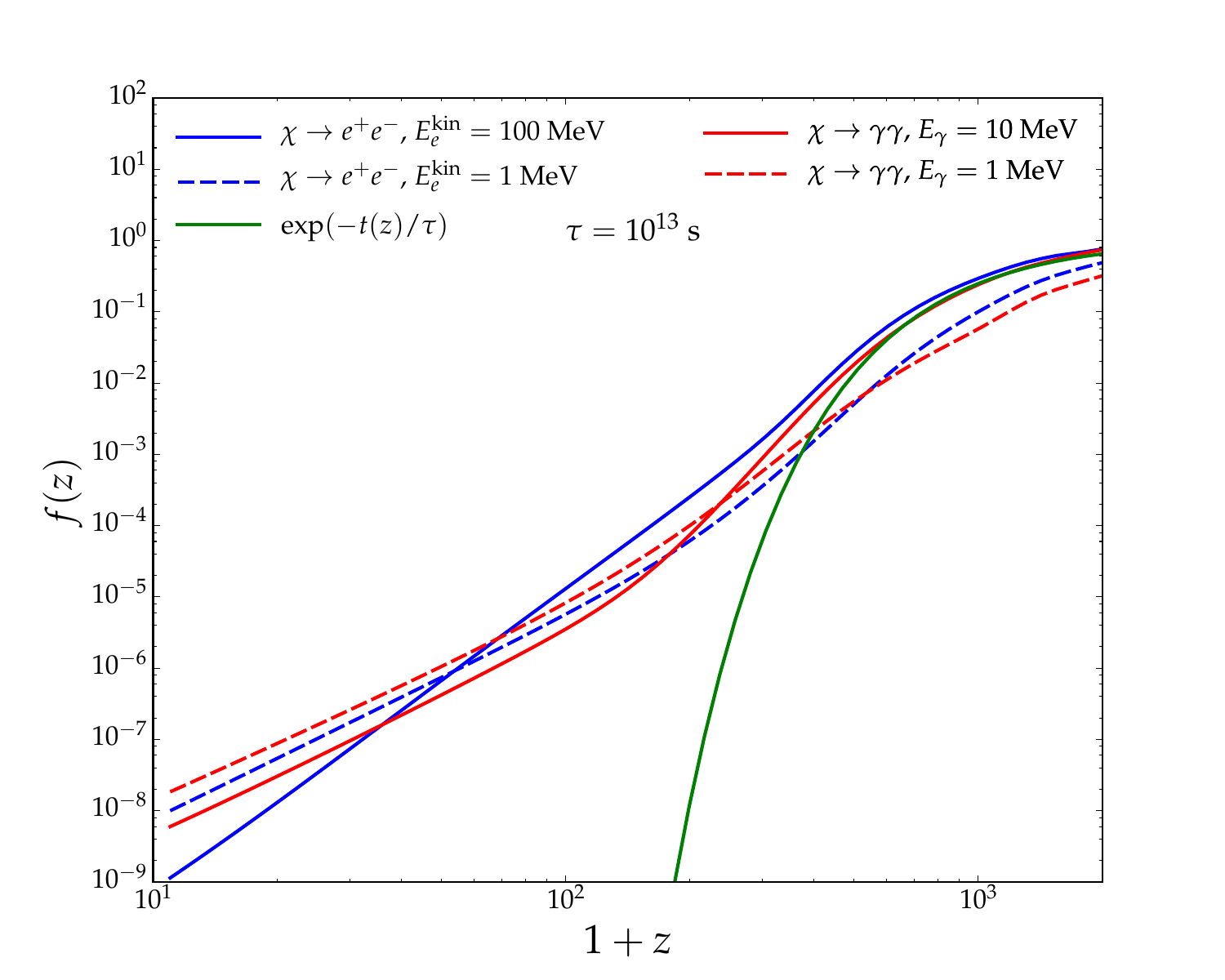}\includegraphics[scale=0.3]{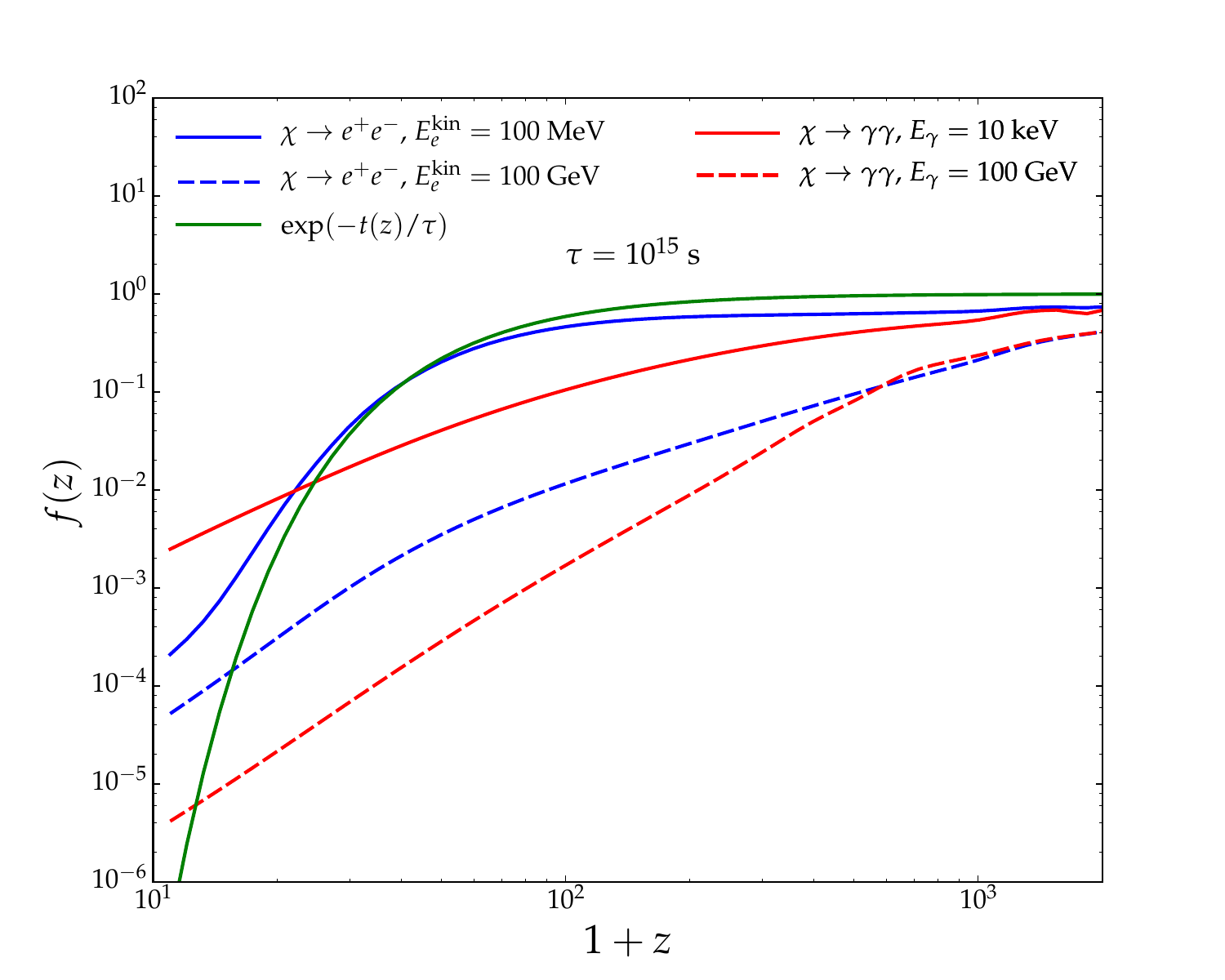}\\
\includegraphics[scale=0.3]{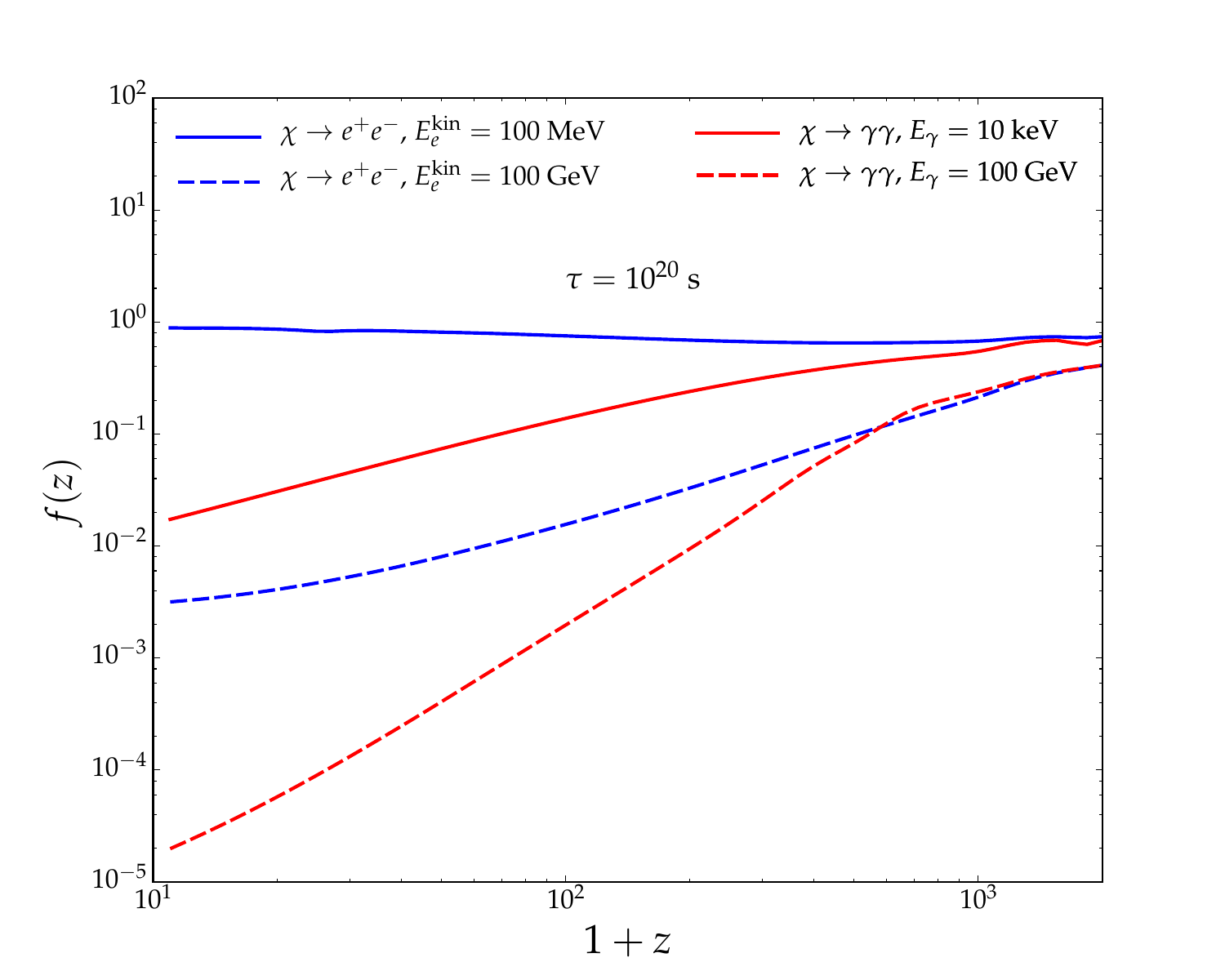}
\caption{\label{Fig:f_z_functions}$f(z)$ functions for particles decaying into electrons and photons, for several injected (kinetic) energies and for three different lifetimes. In the top panels,
the simple exponential decay law for the same lifetime is also reported, for comparison.}
\end{figure}

The $f(z)$ functions are usually obtained in terms of the transfer functions $T_i(E,z,z')$, describing what fraction of the initial energy $E$ of a particle $i$  (in practice either $e^\pm$ or $\gamma$'s) injected at $z'$ is deposited at $z$.
The transfer functions are then simply convoluted with the energy spectrum of each particle $i$, $dN_i/dE$, integrated in time for $z'>z$, hence accounting for the time evolution of the density of the decaying particles, and finally summed over the species, i.e. according to the formula: 
\begin{equation}\label{eq:f_z}
f(z)=H(z)\frac{\sum_\ell\int\frac{d\ln(1+z')}{H(z')}e^{(-\Gamma\,t(z'))}\int T^{(\ell)}(z',z,E)E\frac{dN^{(\ell)}}{dE}\big|_{\rm inj}dE}{\sum_\ell\int E\frac{dN^{(\ell)}}{dE}\big|_{\rm inj}dE}\,.
\end{equation}
The transfer functions have been computed in Refs.~\cite{Slatyer09,Slatyer12} and updated recently in Refs.~\cite{Slatyer15-1,Slatyer15-2}. The main update of latter references over the former is to take into account low energy photons (below 10.2 eV and therefore unable to interact) produced {\em during} the cooling of the high energy particles. Technically, Refs.~\cite{Slatyer15-1,Slatyer15-2} give the transfer functions {\em per channel} for a given ionization history (\texttt{Recfast}-like, without reionization). We cannot make use of the transfer function per channel directly since a large part of the parameter space we are dealing with consists in modification of the reionization history, for which $x_e\simeq 0.1-1$. However, we can use what the authors of the above mentioned references call {\it simplified scheme with 3 keV prescription}, i.e. we rely on the factorization introduced in Eq.~(\ref{eq:factorization}) and that gives very good result for energy injected above $\sim$ 10 MeV. Our treatment corresponds thus to state-of-the art methods of recent literature, although for specific applications {\it ad hoc} calculations might still be needed.  In the following, we will limit ourselves to the injection of  $e^\pm$ and $\gamma$ with a monochromatic energy spectrum. 
This is not necessarily unphysical, since there are many models with a dominant e.m. two body final state. The $\gamma\, \nu$ mode is for instance relevant to
unstable gravitinos with mass below the $W$ gauge boson one; it is also the only ``visible'' decay of a sterile neutrino lighter than 1 MeV. The $\gamma\,\gamma$  final state is the paradigmatic decay channel of the well-known axions/axionlike particles, but it also applies to other cases: One example is a light scalar linearly coupled to matter through the trace of the standard energy-momentum tensor, proposed e.g. as a dark matter candidate from $R^2$ gravity in~\cite{Cembranos:2008gj}. In the same model, if the scalar is heavier than 1 MeV, the $e^\pm$ mode we consider below is the phenomenologically dominant one. Our choice is however mostly dictated by simplicity, since in most decay models a continuous spectrum of e.m. particles is emitted: while our formalism can be applied to any final state with a generic $e^\pm$ and $\gamma$ energy distribution (see Eq.~(\ref{eq:f_z})), it would necessarily be more model-dependent and time-consuming to compute the actual bounds in those cases. Nonetheless, the results obtained in the following are indicative also of cases with a more complicated energy distribution for the final state particles: Since the energy deposition efficiency is typically a smooth function of the particle energy, the energy distributions of the daughter particles can be fairly approximated by replacing them with the deterministic average value, and just correcting for the average energy fractions in e.m. particles. This approximation can be suitably encoded in the parameter $\Xi$ previously introduced. For instance, consider the decay mode $X\to \nu e^+e^-$: if we denote with $x$ the average energy fraction carried away by the neutrino, often a fair proxy of the bound can be obtained from the constraint on the decay of a particle $Y$ into a monochromatic final state, $Y\to e^+e^-$, provided one adopts $m_Y=(1-x)m_X$, and downscales the parameter $\Xi$ by a factor $(1-x)$.
In sec.~\ref{sec:neutrinos}, we will apply this approximate treatment to the specific case of a $\sim 130$ MeV sterile neutrinos whose dominant e.m. decay mode is $\nu_s\to \nu\, e^+e^-$.
In Fig.~\ref{Fig:f_z_functions} we plot the functions $f(z)$ for decays into electrons and photons, for several injected (kinetic) energies (chosen to bracket the energy deposition efficiency) and for three lifetimes ($10^{13}$ s, $10^{15}$ s, and $10^{20}$ s). For comparison, in the two cases for which the lifetime is shorter than the age of the universe, we also plot the decay law $\exp(-t(z)/\tau)$ (green curve, with rapid drop at low $z$) to illustrate its difference with $f(z)$, due to peculiar effects of energy deposition. Obviously, unless the lifetime is very long, $f(z)$ drops dramatically with decreasing $z$ due to the exponential decay factor (top panels). However, even for very long lifetimes such that $\exp(-t(z)/\tau)\sim 1$, $f(z)$ can have a substantial evolution caused by significant changes in the efficiency of energy transfer for particles in a given energy range (bottom panel for the 100 GeV case).

In order to speed up the evaluation of constraints involving different particle physics parameters, such  as different masses or final state channels, further approximations are often introduced.
A very popular one is the effective {\it on-the-spot} approximation. In the context of decaying particles, this approach relies on two assumptions: first, that energy deposition happens at the same redshift as energy injection ($T^{(\ell)}(z',z,E) \propto \delta(z'-z)$ in Eq.~(\ref{eq:f_z}));  second, that the energy deposition efficiency is constant over time up to a factor $e^{-\Gamma t}$, which amounts in assuming that the function $f(z)$ is simply of the form 
\begin{equation}\label{eq:f_z_approx}
f(z)=f_{\rm eff}e^{-\Gamma\,t(z)},
\end{equation}
where the factor $f_{\rm eff}$ only depends on the mass $m_{\chi}$ and on the e.m. branching ratios (i.e. decay channel(s)).
One can think of different Ans\"atze for relating this effective coefficient to the particle properties.

In section~\ref{sec:res}, we will compute bounds on the e.m. decay model without such an approximation. 
Then we will use the beyond-on-the-spot results to physically motivate one particular ansatz for  $f_{\rm eff}$. Finally we will discuss the accuracy of the effective on-the-spot approximation with this ansatz. We can already anticipate here that we will define $f_{\rm eff}$ as the value of the beyond-on-the-spot function $f(z) e^{\Gamma t(z)}$ evaluated at $z\simeq300$. We will see that this gives approximate bounds correct at the $20\%$ level for lifetimes $> 10^{14}\,$s, despite the fact that the exact impact of the decay on $x_e$, for instance, might be poorly described. For shorter lifetimes, no easy criterion emerges.

Let us conclude this section by a technical note. We have modified the {\sf Recfast} \cite{Seager:1999bc} v1.5 routine of the Boltzmann code {\sf CLASS} \cite{Lesgourgues:2011re,Blas:2011rf} v2.5 following recommendations of Refs.~\cite{Chluba:2015lpa,Oldengott:2016yjc}, namely we evaluate the photon ionization coefficient using the photon temperature instead of the electron one. However, we  neglect collisional heating, since this will never be relevant for the level of energy injection not excluded by the CMB constraints: for all the allowed cases of interest, $\TM$  never increases above $10^4\,$K.

\subsection{Effects of electromagnetic decays on the ionization history and the CMB power spectra}

In this section, we describe the impact of the exotic particle decay on the ionization history and on the CMB TT and EE angular power spectra as a function of the lifetime $\Gamma^{-1}$. We show some TT and EE spectra for models whose basic cosmological parameters $\{\theta_s, \omega_b, \omega_{DM}, A_s, n_s, z_{\rm reio}\}$ have been fixed to Planck 2016 TTTEEE+SIMlow best fit \cite{Aghanim:2016yuo}. We illustrate the impact of energy injection for the specific case of a decay into $e^\pm$, but very similar conclusions can be reached for photons and therefore for any e.m. decay product. For each lifetime, we take two  different injected energies, that were chosen to roughly bracket the possible energy deposition efficiencies. A general comment is that, for a given lifetime, different energy deposition histories might lead to slightly different effects on the $C_\ell$'s, giving thereby a potential handle to identify the decay channel and the mass of the decaying particle. Hence, in case of detection, there would be a possibility to pin down the particle physics scenario causing the signal, rather than just measuring the lifetime and abundance of the mother particle. A more careful ``case-by-case'' analysis for interesting scenarios would be needed to further support this statement. 

We wish to test already in this section the accuracy of the on-the-spot approximation,
used in most of past literature, but with our updated ansatz for $f_{\rm eff}$ that will be motivated in section~\ref{sec:res}.
As already hinted above, this consists in replacing $f(z)$ by $f(z=300)e^{-\Gamma [t(z)-t(300)]}$.

\begin{figure}[t!]
\begin{center}
\includegraphics[scale=0.28]{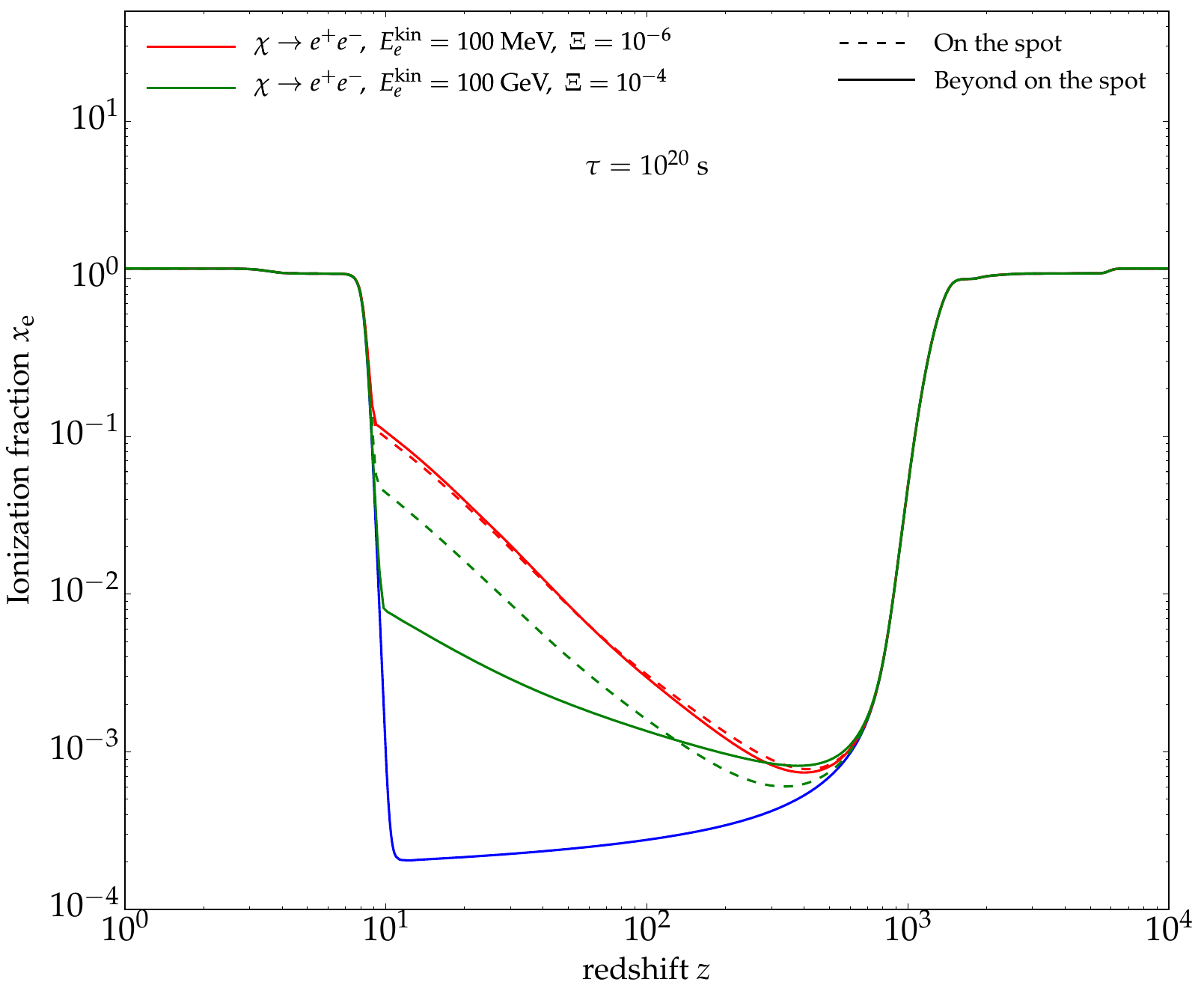}
\includegraphics[scale=0.28]{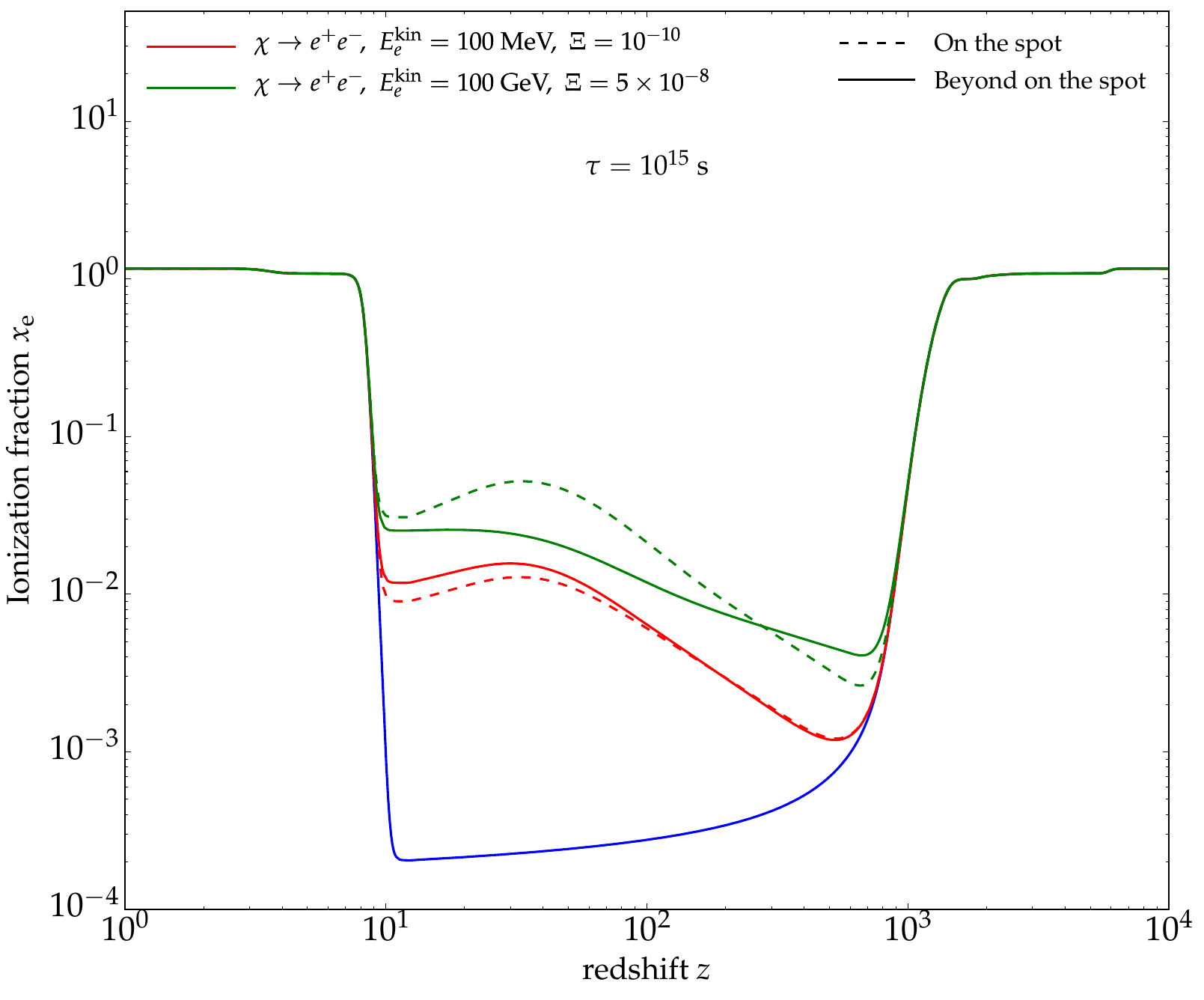}\\
\includegraphics[scale=0.28]{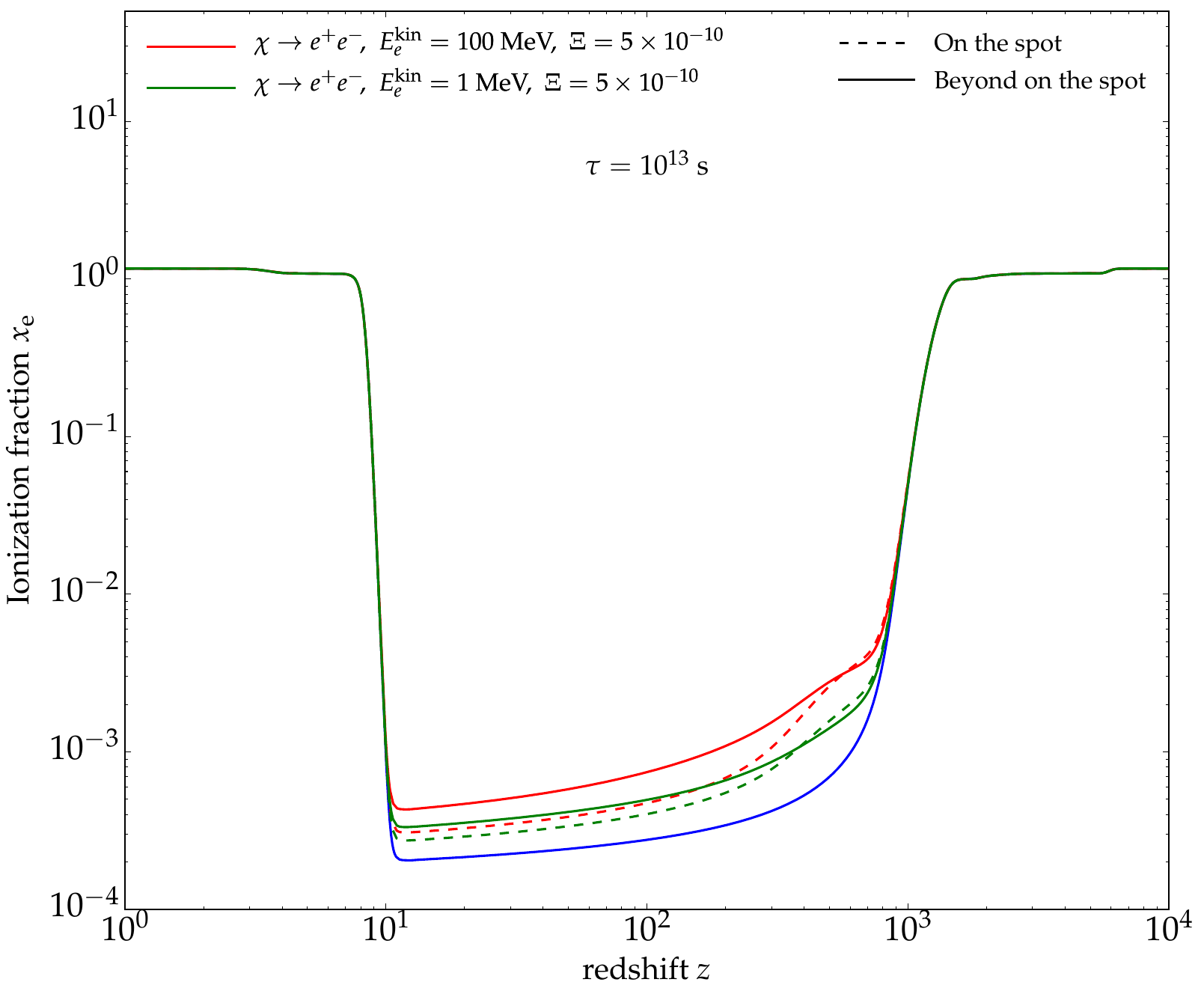}
\includegraphics[scale=0.28]{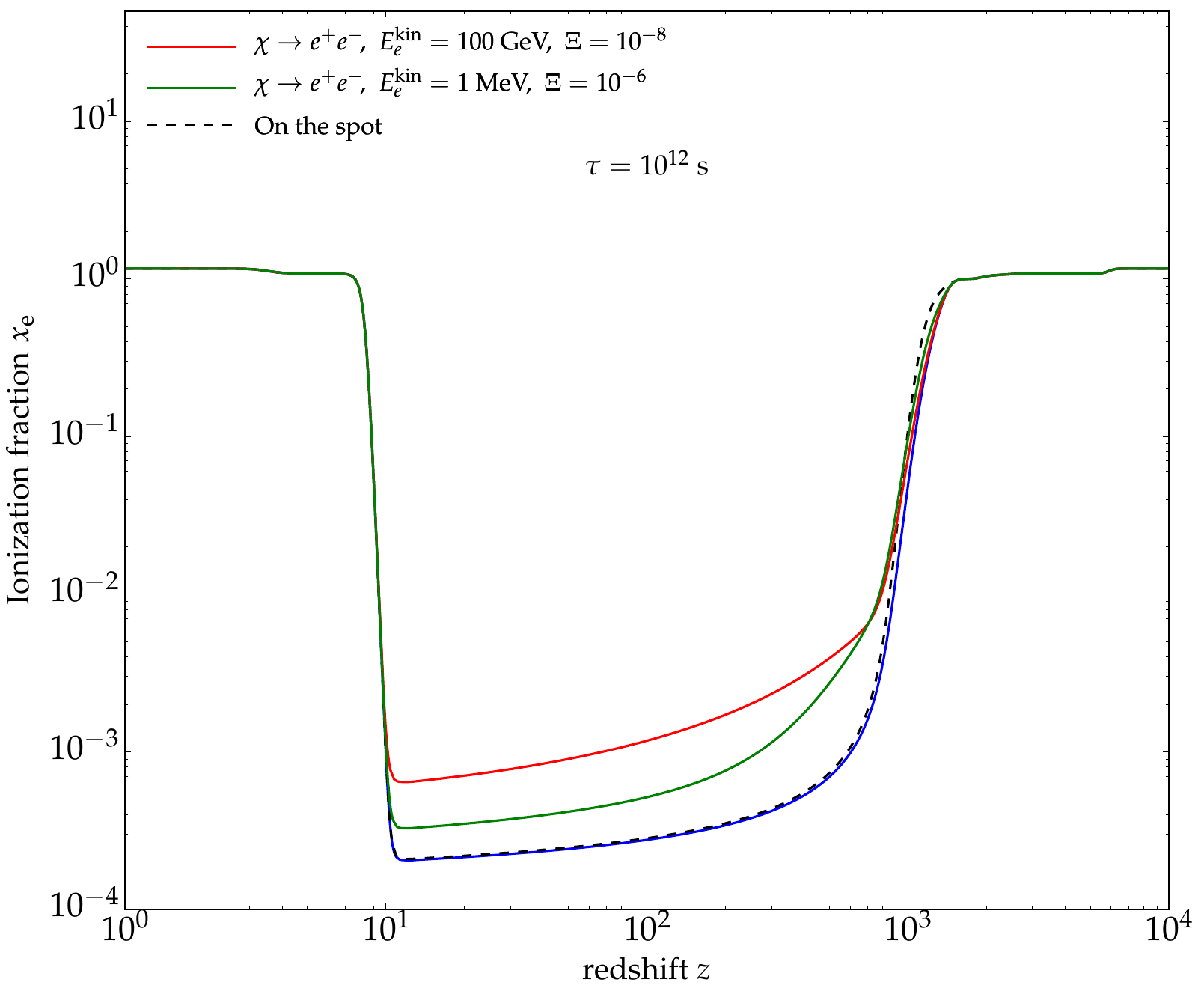}
\caption{\label{Fig:xe_decay} Comparison of the on-the-spot and beyond on-the-spot treatment on the ionization history. We assume a decaying DM model $\chi\to e^+e^-$, with $E^{\rm kin}_e = 100$ MeV, 100 GeV with lifetimes $\Gamma^{-1} = 10^{20}$ s  and 10$^{15}$ s (top-left and top-right panels respectively),  $E^{\rm kin}_e = 1$ MeV, 100 GeV with lifetime $\tau = 10^{13}$ s  bottom-left panel) and $E^{\rm kin}_e = 1$ MeV, 100 MeV with lifetime $\Gamma^{-1} = 10^{12}$ s (bottom-right panel). The blue curves on each plot represent result in the Planck 2016 $\Lambda$CDM model.}
\end{center}

\end{figure}

\begin{figure}[t!]
\begin{center}
\includegraphics[scale=0.33]{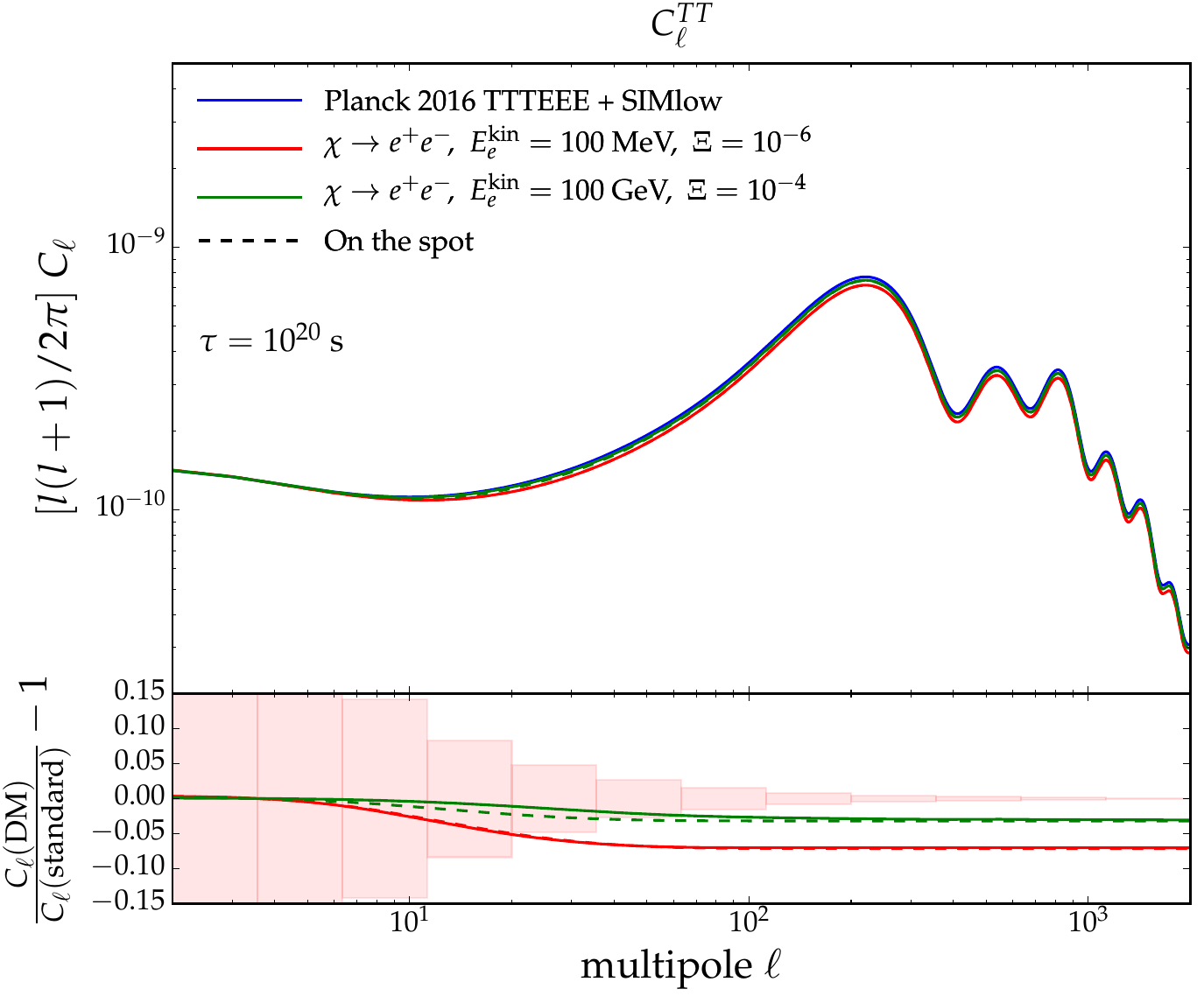}
\includegraphics[scale=0.33]{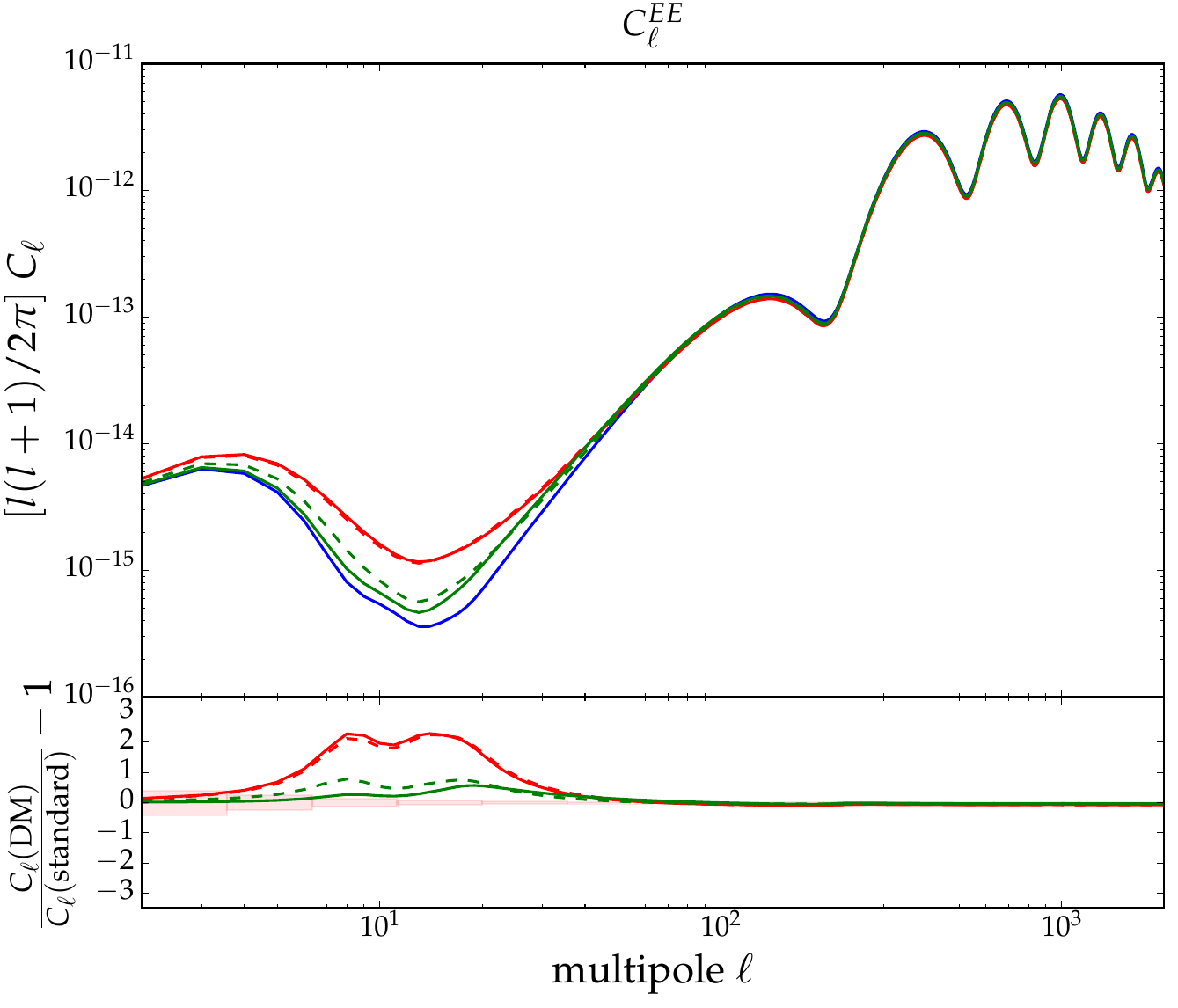}\\
\includegraphics[scale=0.33]{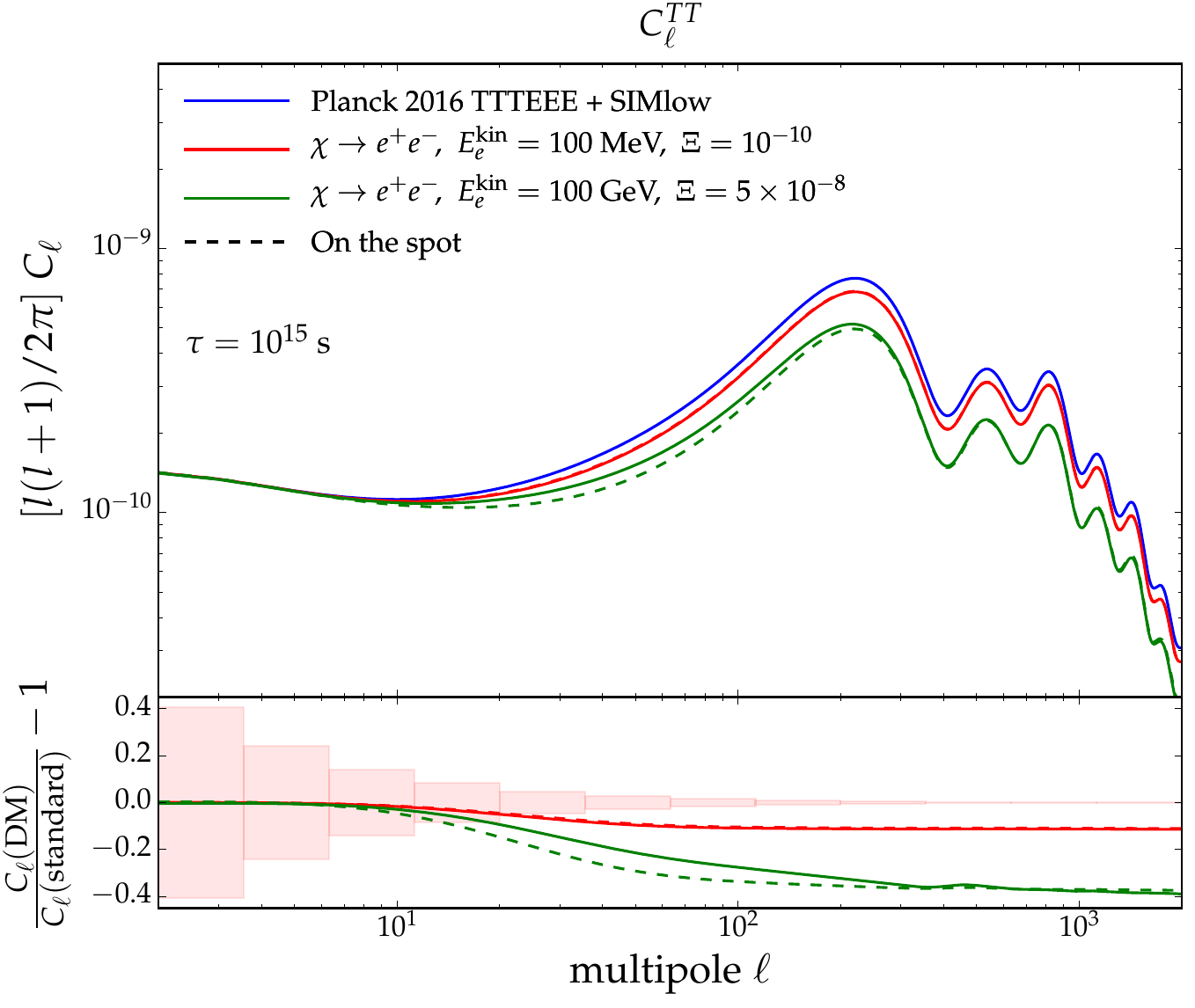}
\includegraphics[scale=0.33]{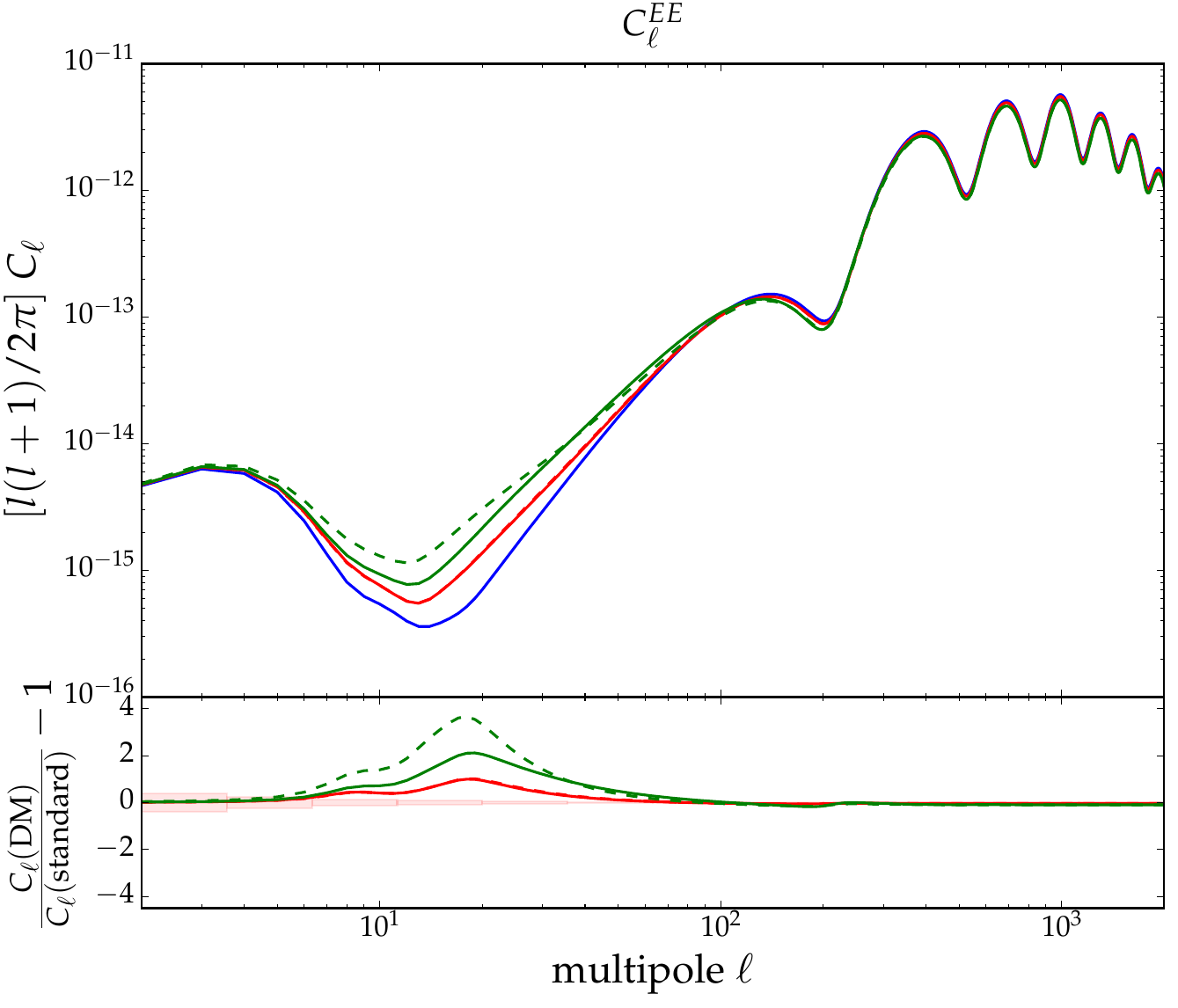}\\
\caption{\label{Fig:Cl_LongLived} Comparison of the on-the-spot and beyond on-the-spot treatment on the lensed temperature and E-mode polarization power spectra, as well as their residuals.  We assume a decaying DM model $\chi\to e^+e^-$, with $E^{\rm kin}_e
= 100$ MeV, 100 GeV with lifetimes $\Gamma^{-1} = 10^{20}$ s (top panels) and 10$^{15}$ s (bottom panels). }
\end{center}

\end{figure}

\begin{figure}[t!]
\begin{center}

\includegraphics[scale=0.33]{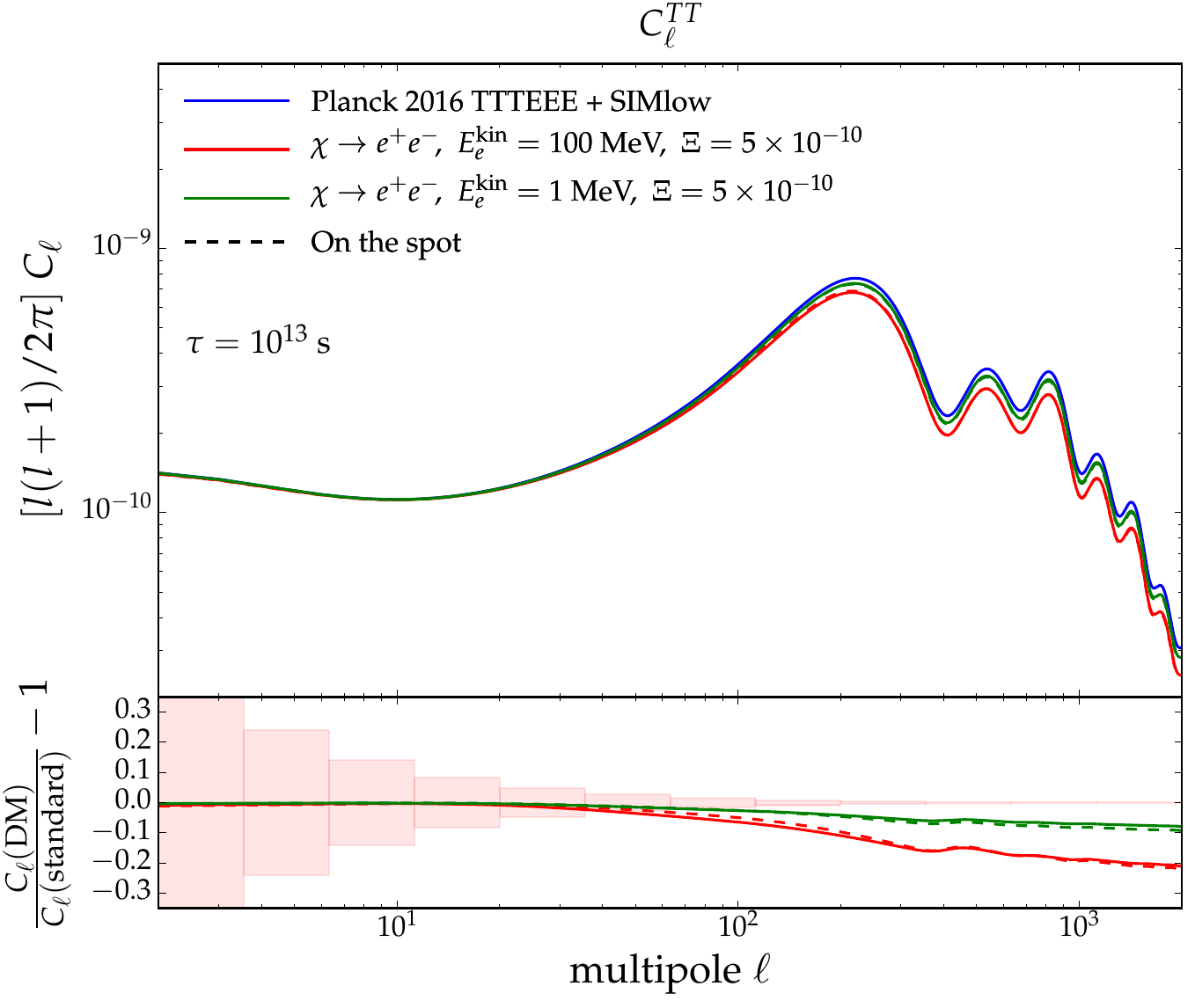}
\includegraphics[scale=0.33]{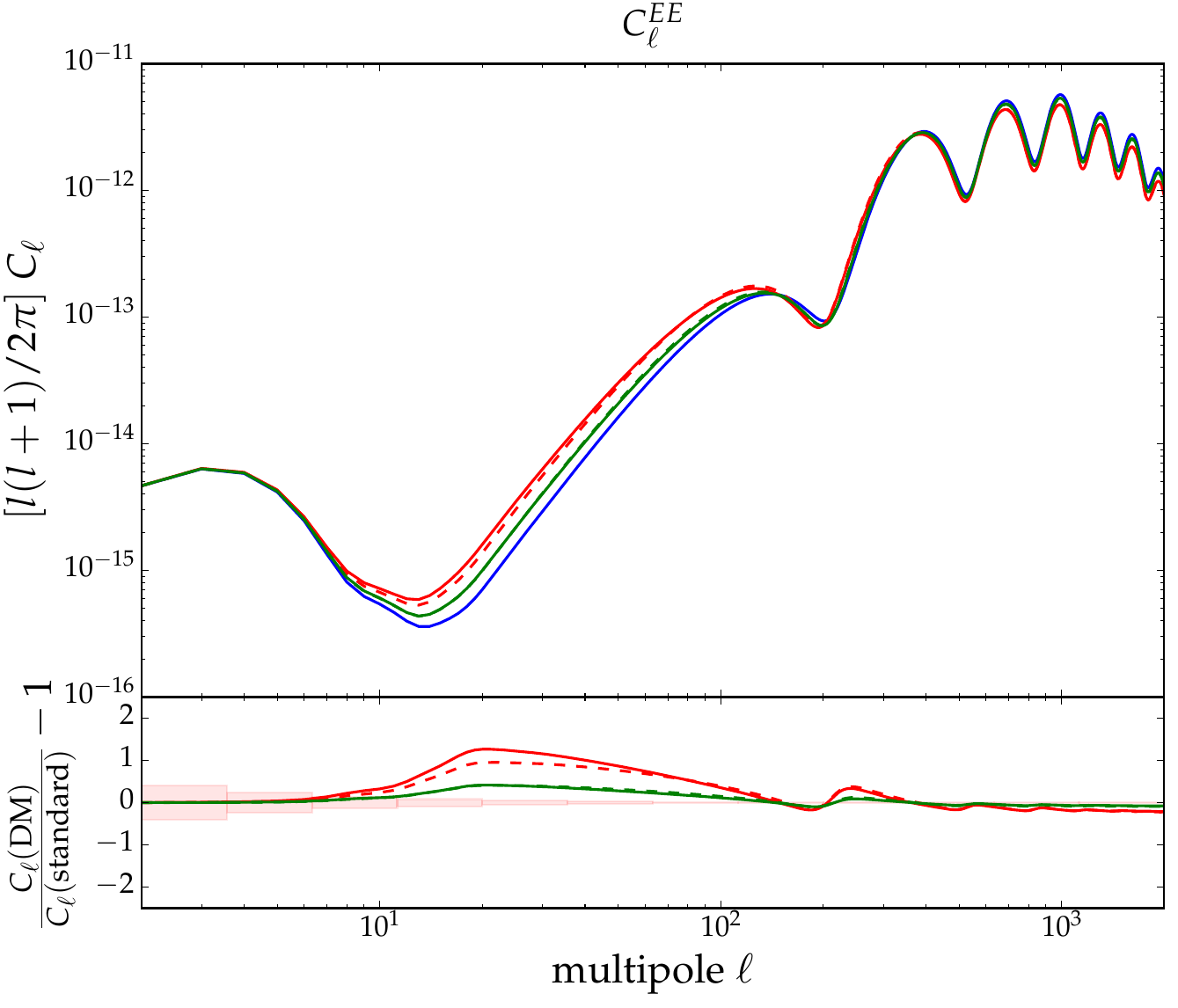}\\
\includegraphics[scale=0.33]{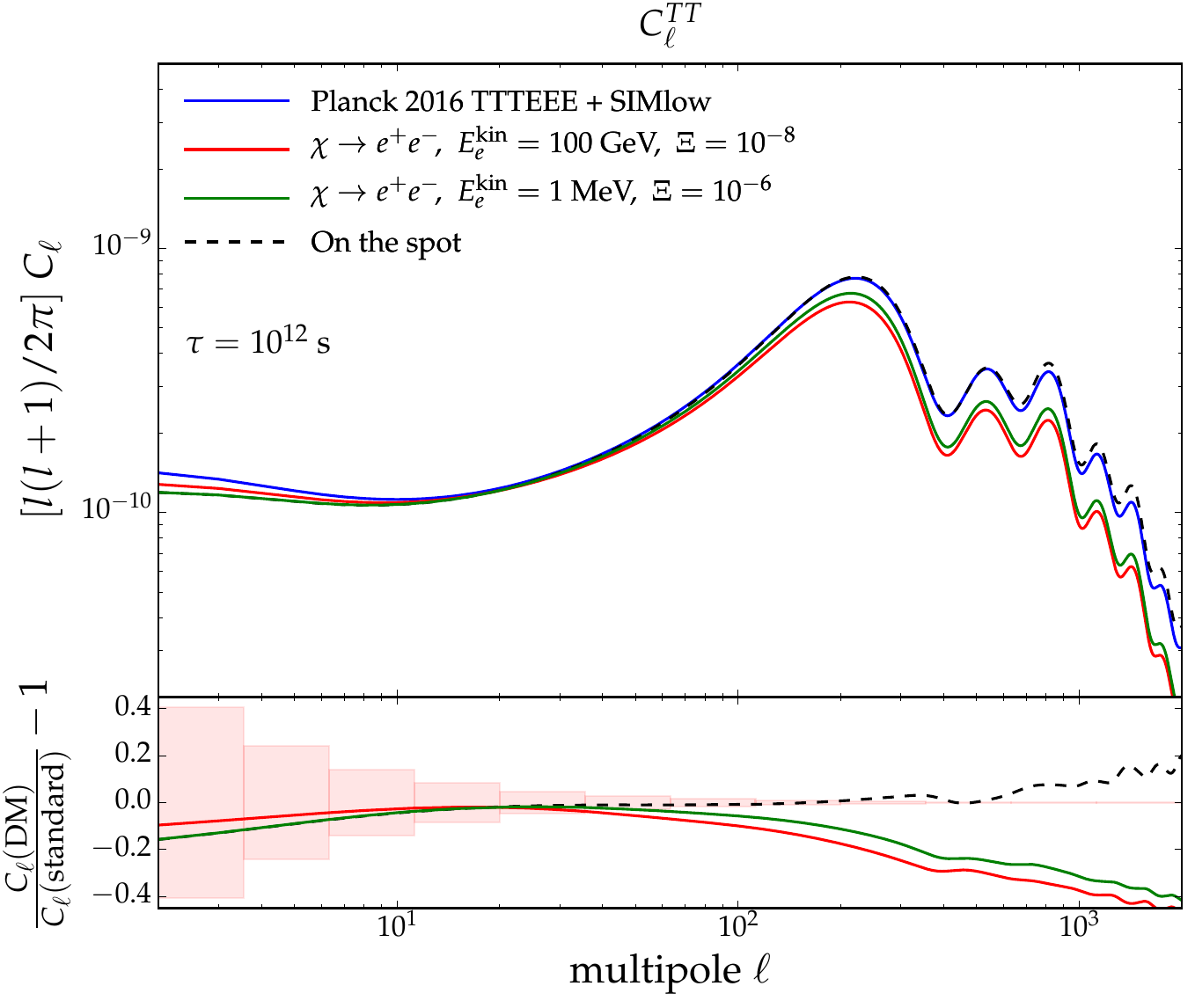}
\includegraphics[scale=0.33]{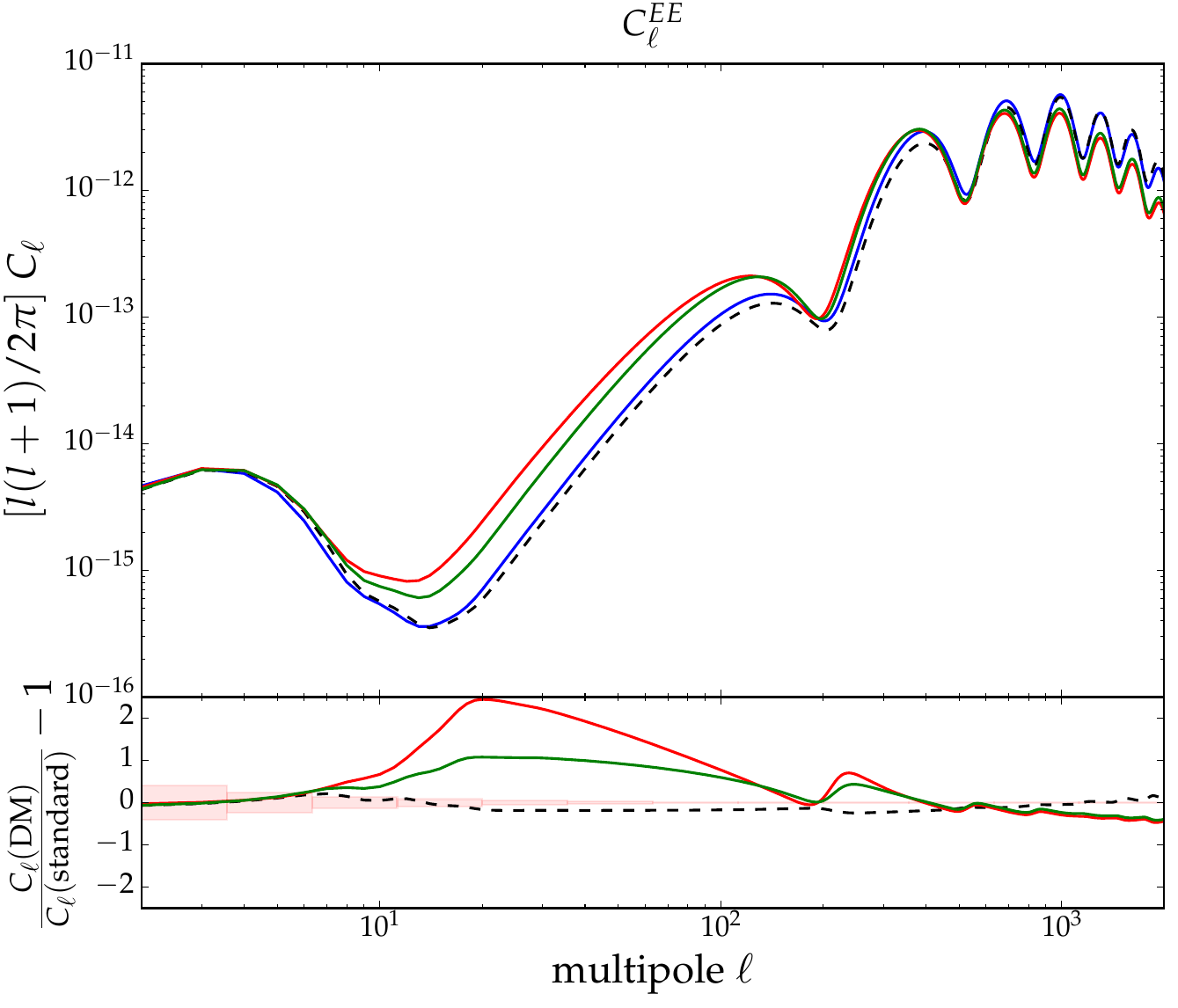}\\
\caption{\label{Fig:Cl_ShortLived}
As in Fig.~\ref{Fig:Cl_LongLived} for decaying DM $\chi\to e^+e^-$ with lifetime $\Gamma^{-1} = 10^{13}$ s and  injected energies  $E_e =$ 1 MeV and 100 MeV (top panels), as well as lifetime $\Gamma^{-1} = 10^{12}$ s and injected energies $E_e =$  1 MeV and 100 GeV (bottom panels).}
\end{center}
\end{figure}

Let us consider separately different possible orders of magnitude for the particle lifetime:
\begin{itemize}
\item[\textbullet] as found in previous literature (see e.g. Refs.~\cite{Diamanti:2013bia,Oldengott:2016yjc, Chen}), for a lifetime bigger than the age of the universe, $\tau_u \simeq 10^{18}$ s, the main effect of the particle decay is to initiate slowly the reionization of the IGM at high redshift $z>100$ (Fig.~\ref{Fig:xe_decay}, top-left panel). Looking at the CMB power spectra plotted on Fig.~\ref{Fig:Cl_LongLived}--upper panels, the decay of long-lived particles imply a step-like suppression of the TT and EE power spectra,
plus a much stronger and wider reionization bump in the case of the EE spectrum.\\
In the case of long-lived particles, most of the effect of the decay on the CMB spectra is well captured by our on-the-spot approximation, as can be seen on the same figure, red curve. The exponential factor is very close to unity even at $z=0$, and the effect of the decay can be fully described by the unique combination of parameters 
\begin{equation}
\xi  \equiv f_{\rm eff}\,\Xi\,\Gamma
\label{eq:xi}
\end{equation}
which has the dimension of a rate.\\
The CMB temperature spectrum probes the reionization history after recombination mostly through the integrated quantity $\tau_{\rm reio}$. Hence, the scenarios presented in the figure (all sharing the same $z_{\rm reio}$) give identical temperature spectra up to corrections below the percent level, despite of differences in the ionization history in the range $z_{\rm reio} \leq z \leq 10^3$. However, the EE spectra is more sensitive to the ionization history and we can see variations of the order of $50\%$ at $\ell < 10$ for the 100 GeV case.
 Note that the quality of the on-the-spot approximation decreases when the  injected energy increases and the energy deposition efficiency goes down.

\item[\textbullet] For $\Gamma^{-1} \gtrsim 10^{14}$ s (Fig.~\ref{Fig:xe_decay}, top-right panel), a bump in the free electron fraction appears, localized around the time of the decay. However, the recombination history is not impacted. The enhancement of the optical depth integrated up to the surface of last scattering leads to a larger step-like suppression of the TT and EE spectrum on small angular scales (Fig.~\ref{Fig:Cl_LongLived}, bottom panels). The slightly increased probability of photons to re-scatter  at intermediate redshifts (between recombination and reionization) generates extra polarization, and leads to a characteristic bump in the EE spectrum, peaking on smaller angular scales than the usual reionization bump (around $\ell \simeq 20$ in the examples displayed in Fig.~\ref{Fig:Cl_LongLived}--bottom-right panel, instead of $\ell=3$ for reionization). This could be a rather unique signature of the peculiar reionization history in these models, and the measurement of the bump location in multipole space would give a direct indication on the DM lifetime.\\
Comparing now the results of the accurate treatment with the on-the-spot approximate version, we find that the latter works well for particles decaying with substantial energy deposition efficiency (here for 100 MeV), but not for high  injected energies (100 GeV in the example of Fig.~\ref{Fig:Cl_LongLived}, bottom panel, green curve). 
At 100~GeV, the error is as large as $10\%$ around $\ell\simeq 50$ in the TT spectrum, and $100\%$ around $\ell\simeq 20$ in the EE spectrum. 

\item Finally, for very short lifetimes $\Gamma^{-1}\lesssim 10^{13}$ s  (Fig.~\ref{Fig:xe_decay}--bottom-left panel), the decay starts to modify also the recombination era, eventually delaying it. The most visible consequence in the CMB temperature spectra is a stronger damping tail (Fig.~\ref{Fig:Cl_ShortLived}, top panels). This comes from the enhanced Silk damping effect caused by a larger width of the last scattering surface. The bump coming from extra re-scattering and polarization is also visible, but is less sharp than in the previous case. For a lifetime of $\Gamma^{-1}\simeq 10^{12}$ s, some more peculiar patterns appear even on large angular scales (Fig.~\ref{Fig:Cl_ShortLived}--bottom panels). The reason is that we are running {\sf CLASS} with a fixed angular sound horizon scale $\theta_s$. In these models, recombination is delayed significantly, and the sound horizon at recombination is larger. The code automatically adapts the  angular diameter distance to the last scattering surface by {\em decreasing} $H_0$, to maintain the same $\theta_s$. Hence the late-time expansion history is modified and this results in a smaller ``late integrated Sachs-Wolfe'' effect on large scales.\\
Fig.~\ref{Fig:xe_decay} shows again the results obtained using the on-the-spot approximation with our Ansatz $f_{\rm eff} = f(300)e^{\Gamma t(300)}$. In the short lifetime limit, we see that this approach does not capture the effects of the decay, and would not lead to reliable constraints, which is not surprising because most particles decay {\it before} $z=300$. Hence one would need to adapt the ansatz for each lifetime. In the specific case of $\Gamma^{-1}= 10^{13}$ s, we have been able to find by trial and error that the ansatz $f_{\rm eff}=f(z_{\rm decay})e^{\Gamma t(z_{\rm decay})}$ (with $t(z_{\rm decay})=\Gamma^{-1}$ by definition) leads to effects similar to the beyond-on-the-spot result. 
In that case, the curves are still in reasonable agreement, especially in the TT spectrum where the agreement reaches the percent level, while differences in the EE spectrum are kept below 30$\%$. On the other hand, in the case of $\Gamma^{-1}\sim 10^{12}$ s, the on-the-spot spectra fails even to predict the correct shape for the effect of the decay on the CMB spectra.
For such small lifetimes, the lack of a meaningful physical criterion to even define an on-the-spot approximation makes any simplification attempt hazardous. A more refined search for a phenomenological criterion, e.g. via a principal component analysis as done in the DM annihilations case \cite{Slatyer12}, would probably be useless. In that case, in order to get reliable constraints, one has to do MCMC scans for each lifetime, as we shall perform in the following. 
\end{itemize}


In summary, we have illustrated the discrepancy between the on-the-spot approximation and the more accurate beyond-on-the-spot treatment, for different lifetimes, injected energies and decay products. These discrepancies are more obvious in $x_e(z)$ than in the $C_\ell$'s. We found that {\em for lifetimes $\geq 10^{13}$ s the approximation can reach the $\%$ level agreement if an appropriate criterion is chosen}, like our ansatz $f_{\rm eff}=f(300) e^{\Gamma t(300)}$ that we will further justify in the next section. However, the error increases when 
the energy deposition efficiency is very different from an exponentially decaying function: this is the case for large  injected energies and thus  high DM masses.
The error can largely exceed the 10-20\% accuracy expected from a state-of-the-art treatment in terms of tabulated transfer functions. Furthermore, for shorter lifetimes ---although we did not investigate deeper in this issue---no simple criterion emerges which is at the same time accurate and  universal enough to be of practical use. Hence, the on-the-spot approximation should only be used for estimating the order of magnitude of the CMB bounds in a restricted window in parameter space. In the next section we will work with the beyond-on-the-spot approach,  in order to get results accurate at the 10-20\% level.

\section{Results: Summary of constraints and comparison with other probes}\label{sec:OtherConstraints}
\subsection{Methodology}
We perform our study of CMB anisotropy constraints with Monte Carlo Markov chains, using the public code {\sc Monte Python} \cite{Audren12} and the Metropolis Hasting algorithm. Beyond the on-the-spot approximation and for each decay channel, the sector of e.m. decaying particles can be described by three independent parameters, e.g. the particle mass, lifetime and effective energy density parameter $\Xi$ defined in section~\ref{sec:standard_eq}. For a better efficiency of MCMC runs, our strategy consists in scanning a grid of values for the lifetime $\tau \equiv \Gamma^{-1}$ in the range [$10^{12}$s; $10^{26}$s], and for values of the mass $m$ leading to byproducts with (kinetic) energies in the range [10 MeV; 1 TeV]. For each set of ($\tau, m$) values, we perform a fit to the data with flat priors on the following set of parameters:
$$
\Lambda{\rm CDM}\equiv\{\omega_b,\theta_s,A_s,n_s,\tau_{\rm reio},\omega_{\rm DM}\}+\Xi\,.
$$
We use a Choleski decomposition to handle the large number of nuisance parameters in the Planck likelihood \cite{Lewis:2013hha}. We consider chains to be converged when the Gelman-Rubin \cite{Gelman:1992zz} criterium gives $R -1<0.05$. 
We use the Planck 2015 high-$\ell$ TT,TE,EE likelihood, a prior on $\tau_{\rm reio}$ taken from the Planck 2016 results based on the new SimLow likelihood \cite{Aghanim:2016yuo}, and the Planck 2015 lensing likelihood.

Since the effect of a long-lived decay is similar to modifications of the reionization history, one might expect constraints to depend on the way in which reionization itself is modelled.  Thus we first perform a couple of runs in order to assess the impact of assuming either instantaneous (``camb-like'') reionization\footnote{In the instantaneous reionization, the free electron fraction is given at low-$z$ by $x_e(z) = \frac{f}{2}\big[1+\tanh(\frac{y-y_{\rm re}}{\Delta y})\big]$ with $f = 1 + n_{\rm He}/n_{\rm H}$, $y = (1 + z)^{3/2}$ and $\Delta y = 3(1 + z)^{1/2}\Delta z$. The reionization is therefore redshift-symmetric, centered around the key parameter $z_{\rm re}$ with a width given by $\Delta z$. }, or the redshift-asymmetric parametrization of \cite{Douspis:2015nca} given by \footnote{Like the authors of Ref.~\cite{Oldengott:2016yjc}, we replaced the argument of the exponent by $-\lambda \frac{(z-z_{\rm p})^3}{(z-z_{\rm p})^2+0.2}$ in order to improve the smoothness of the transition.}
\begin{equation} \label{eq:reio_param_2}
x_e(z) =f \times \left\{ \begin{array}{cl}
& \frac{1-Q_{\rm p}}{(1+z_{\rm p})^3-1}\big((1+z_{\rm p})^3-(1+z)^3\big)+Q_{\rm p}\:\:\;\textrm{for } z < z_{\rm p} \\
& Q_{\rm p}\exp\big(-\lambda (z-z_{\rm p})\big) \:\:\;\textrm{for } z \geq z_{\rm p}.
\end{array}\right.
\end{equation} 
The parameters of the second model have been estimated to be close to $z_{\rm p} = 6.1$, $Q_{\rm p}\equiv Q_{\rm HII} (z_{\rm p})  = 0.99986$ and $\lambda = 0.73$ by direct observations of the ionized hydrogen fraction $Q_{\rm HII} (z)$. 
Similarly to Ref.~\cite{Oldengott:2016yjc}, we fix $z_{\rm p}$ and $Q_{\rm p}$ to their best-fit values, and we let the evolution rate $\lambda$ free to vary in order to cover a large range of possible ionization histories. Note that we could have used instead the power-law asymmetric reionization suggested by Ref.~\cite{Adam:2016hgk}. In fact, we have checked explicitly that differences between the two parameterizations are not visible at the current sensitivity level, and most probably even for future CMB experiments. 

For these test runs, we use for simplicity the on-the-spot approximation, and we are only interested in long-lived particles. Hence the only relevant parameter describing the e.m. decay sector is the combination $\xi$ defined in Eq.~(\ref{eq:xi}).
With camb-like reionization, we obtain $\xi<5.28\times10^{-26}\,{\rm s}^{-1}$ (95$\%$ CL), whereas the use of the redshift-asymmetric parameterization yields $\xi<6.03\times10^{-26}\,{\rm s}^{-1}$ (95$\%$ CL),  which is comparable to the results of Ref.~\cite{Oldengott:2016yjc}. The difference is only at the level of 15$\%$, which is anyway the accuracy level expected for the theoretical calculation of transfer functions in the beyond-on-the-spot approximation, and hence the precision with which we expect to estimate the decay lifetime in general using CMB data. We conclude that assumptions on the reionization history have a marginal impact on lifetime bounds, and from now on we will always rely on the instantaneous camb-like parametrization.\\

\subsection{Comparison of various constraints}
The bounds from CMB angular power spectra are not the only cosmological bounds available. In order to assess their relevance, we compare them with: i) the constraints from CMB {\it spectral} distortions, as bounded by COBE-FIRAS~\cite{Fixsen:1996nj};  ii) the constraints from light nuclei overproduction/destruction with respect to standard BBN prediction. Below, we quickly summarize our treatment of these probes, referring to specialized literature for details.

\subsubsection{CMB spectral distortions from DM decay \label{sec:dist}}
Since the seminal series of papers~\cite{Zeldovich:1969ff, Sunyaev:1970eu, Sunyaev:1970er}, lots of efforts have been paid towards a better treatment of these distortions (see e.g. \cite{Chluba:2011hw} for a recent review), including applications to early decaying massive particles (e.g. \cite{Hu:1993gc,1992NuPhB.373..399E,Chluba:2013wsa,Chluba:2013pya}). It has been found that the distortions are (mostly) of two types, depending on the energy injection time. At very early times, no spectral distortion can survive, since it would simply result in a shift of the blackbody temperature.  Below $z\sim2\times10^6$, the CMB spectrum starts acquiring an effective chemical potential $\mu$, since processes changing the number of photons such as double Compton scattering become inefficient. At later time, the CMB is
mostly sensitive to a modification of the so-called comptonization $y-$parameter quantifying the amount of energy transfer via Compton scattering, which becomes less and less efficient. At first order, one can consider that the transition between the two types of distortions takes place abruptly around $z \simeq 4\times 10^5$, i.e. $\Gamma^{-1} \simeq 5\times10^{10}$s~\footnote{More recently, it has been found that the transition requires a careful numerical computation and may lead to a new type of distortions - the so-called $r$-distortion \cite{Chluba:2011hw}. We will neglect this for simplicity in the present work.}. 
In order to compute the fractional  energy density change of CMB photons, $\Delta\rho_\gamma/\rho_\gamma$, we follow the formalism of \cite{Chluba:2013pya,Chluba:2016bvg}:
\begin{equation}
\frac{\Delta\rho_\gamma}{\rho_\gamma}\approx\bigg[\frac{\Delta\rho_\gamma}{\rho_\gamma}\bigg]_\mu+\bigg[\frac{\Delta\rho_\gamma}{\rho_\gamma}\bigg]_y~,
\end{equation}
where the pure $\mu-$ and pure $y-$distortions can be computed as 
\begin{equation}
\frac{\mu}{1.401}=\bigg[\frac{\Delta\rho_\gamma}{\rho_\gamma}\bigg]_\mu\simeq  \int \mathcal{J}_{\rm bb}\mathcal{J}_{\mu}\frac{1}{\rho_{\gamma}}\,\frac{dE}{dt}\bigg|_{e.m.}\,dt,\qquad \frac{y}{4}=\bigg[\frac{\Delta\rho_\gamma}{\rho_\gamma}\bigg]_y\simeq \int \mathcal{J}_{\rm bb}\mathcal{J}_{y}\frac{1}{\rho_{\gamma}}\,\frac{dE}{dt}\bigg|_{e.m.}\,dt\,.
\label{eqs:mu_y}
\end{equation}
The visibility functions are given by~\cite{Chluba:2013vsa}
\begin{eqnarray}
\mathcal{J}_{\rm bb}(z) \approx\exp[-(z/z_\mu)^{5/2}]\,,\quad
\mathcal{J}_{y}(z)  \approx\bigg[1+\bigg(\frac{1+z}{6\times10^4}\bigg)^{2.58}\bigg]^{-1}\!\!,\quad\mathcal{J}_{\mu}(z)\approx 1-\mathcal{J}_{y}\,.
\end{eqnarray}
In previous equations, $z_\mu= 1.98 \times 10^6(\Omega_bh^2/0.022)^{-2/5}[(1-Y_p/2)/0.88]^{-2/5}$ is the thermalization redshift (due to double-Compton processes) and $Y_p\simeq 0.245$ stands for the primordial mass fraction of Helium-4. These equations can only be trusted until the recombination time, for $z \geq 1000$, but this is sufficient for our purpose.

Indeed, the additional contribution to $y$ coming from low redshifts can be estimated with the standard approximate integral:
\begin{equation}\label{eq:y_standard}
y =\int\frac{k(T_e-T_\gamma)}{m_ec^2}\sigma_t n_e\, c\, {\rm d}t'\,.
\end{equation} 
A long-lived particle could in principle reheat the IGM and contribute to the $y-$distorsions through this integral. 
However, this contribution would remain below the order of $10^{-8}$, well below the current experimental sensitivity, and more importantly, two orders of magnitudes below the $y-$distortions expected to be generated during the reionization era in the standard cosmological scenario.  Hence, in the future, it would be challenging to disentangle exotic contributions to the $y-$parameter generated by a post-recombination particle decay from those produced by standard reionisation. We conclude that CMB spectral distorsions are not relevant for constraining long-lived particles  (at least the global signal).
On the other hand, spectral distortions remain of major interest for particle decaying before recombination, which are loosely constrained by the CMB angular spectra (see also Ref.~\cite{Chluba:2013pya} for a detailed analysis of sensitivity prospects as a function of the particle lifetime). We wish to report bounds in that case, and for that purpose we can neglect contributions to the $\mu-$ and $y-$parameters below $z\leq 1000$. Hence we perform the integrals of equations~(\ref{eqs:mu_y}) down to $z=1000$ only. Our bounds are thus conservative. 

\subsubsection{Constraints from primordial nucleosynthesis \label{sec:bbn}}

BBN has been used since decades to derive constraints on exotic energy injection, in particular from e.m. particle decay as considered here. Energy injection could modify the yields of light nuclei, whose observations match pretty well the predictions of the $\Lambda$CDM model (see e.g. Refs~\cite{Iocco:2008va,Pospelov:2010hj} for reviews). In what follows,
we focus on possible modifications of the ${}^2$H and ${}^3$He abundances, which are known to be at the same time very sensitive and relatively robust probes. We adopt the formalism and use the results reported in Ref.~\cite{Poulin:2015opa}. In that work, it was noted that for sufficiently low energy photons and/or late injection, standard treatments in the literature relying on a universal non-thermal photon spectrum may underestimate the bounds. We neglect this subtely and quote results obtained
with the  universal spectrum, which are robust and conservative. 

The abundance of Lithium-7 could offer a complementary probe of e.m. energy injection in the early universe. However, the relation between its observed `quasi-plateau' abundance in metal-poor halo stars and its cosmological abundance is still disputed. Assuming that the two are equals, it has been recently shown by two of us that a purely e.m. injection of energy may reconcile observations with theoretical predictions~\cite{Poulin:2015woa}  (see also \cite{Salvati:2016jng} for an update). Since this is not the point of the present paper, in which we wish to derive robust bounds on the e.m. decay lifetime, we will not consider Lithium-7 in what follows. 

\begin{figure}[t]
\centering
\includegraphics[scale=0.5]{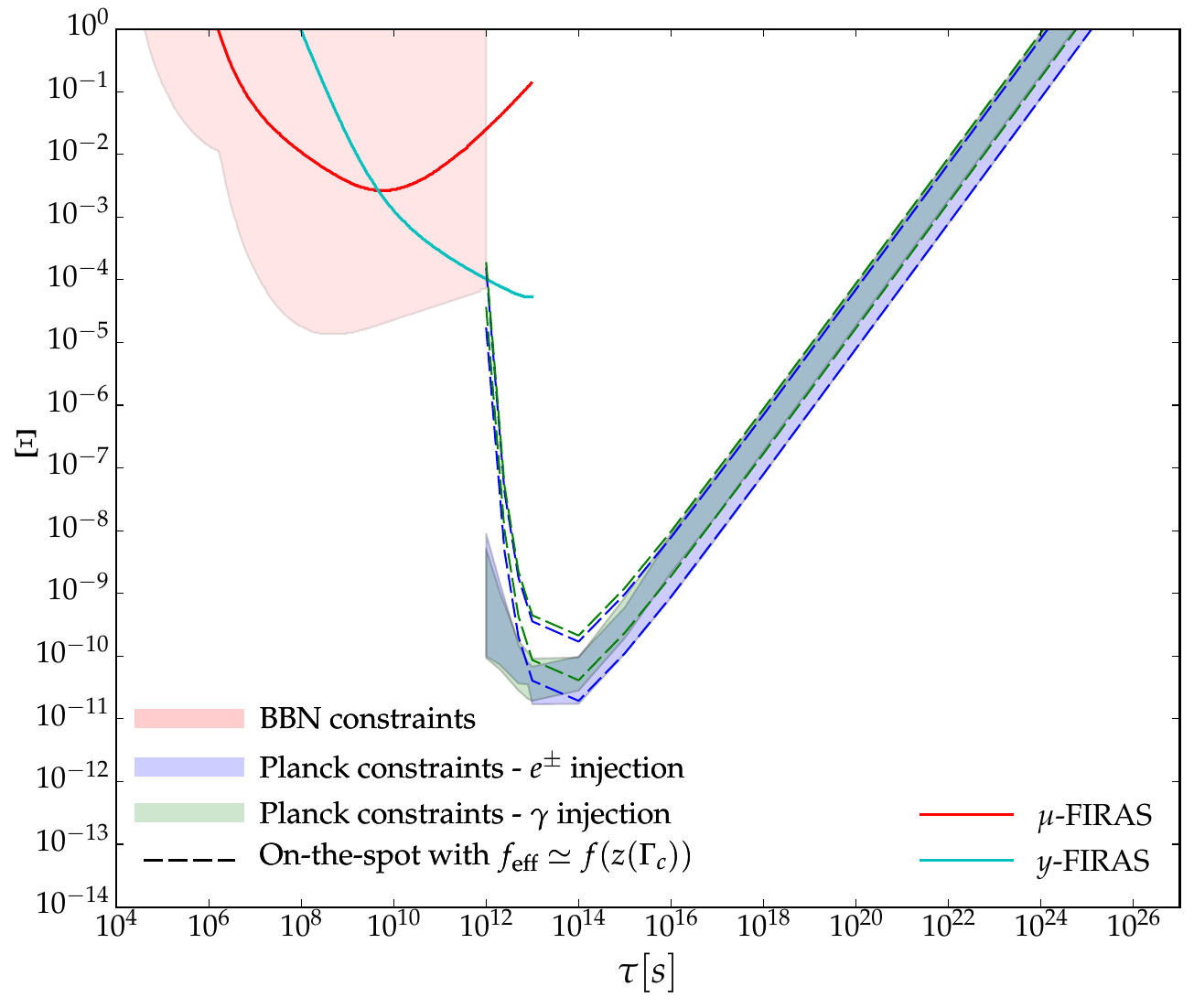}
\caption{\label{Fig:AllConstraints}Cosmological constraints on the effective energy density parameter $\Xi$ of exotic particles with e.m. decay channels. The effective energy density $\Xi$ is normalised to the CDM energy density, and incorporates an efficiency factor (thus $\Xi=1$ means that the particle makes up 100\% of CDM, and that 100\% of the decay energy goes into stable photons and $e^{\pm}$).
We report bounds coming from Big Bang Nucleosynthesis (shaded red area), CMB spectral distortions (full lines) and CMB anisotropies measured by Planck (shaded blue area for $e^\pm$ and shaded green area for $\gamma$; the width of the band is obtained by scanning over the  kinetic energy of the injected particles in the range [10 keV; 1 TeV]). In the case of CMB anisotropies, we compare the use of the full treatment for the energy deposition with the on-the-spot approximation (dashed lines).}
\end{figure}

\subsubsection{Results \label{sec:res}}

We summarise the constraints on massive particles decaying into photons and $e^\pm$  with kinetic energies in [10 MeV; 1 TeV] in Fig.~\ref{Fig:AllConstraints}. 
 In fact, due to the behaviour of injection efficiency with energy, see \cite{Slatyer15-2}, the range of CMB constraints thus obtained also covers the case of smaller injected energies, down to  about 10 keV. Hence, we can safely assume that the band displayed gives a very good estimate of constraints for energies in the range [10 keV; 1 TeV].
For each lifetime, we report lower bounds on the effective energy density parameter $\Xi$, defined in section~\ref{sec:standard_eq}. Given this definition, if the decaying particle accounts for 100\% of cold dark matter, $\Xi$ is expected to be equal to one or at least of order one, unless no sizable amount of $e^{\pm}$ and photons are produced by the decay (e.g. the particle may decay entirely into neutrinos or dark radiation, a case considered separately in Ref.~\cite{Poulin:2016nat}). If at least a small fraction of the injected energy goes into photons, electrons and positrons, a bound $\Xi \ll 1$ implies that the particle contributes to the total dark matter density by a negligible amount.

For each assumed lifetime,  injected energy and decay channel ($\gamma$ or $e^{\pm}$), our {\sc MontePython} run gives a CMB upper bound on $\Xi$ that we report on the figure. We obtain two bands, one for each decay channel, with a width corresponding to extreme assumptions on the injected energy. We also repeat the analysis with the on-the-spot approximation and show the results with dashed lines.

Figure~\ref{Fig:AllConstraints} also shows our constraints coming from $\mu-$ and $y-$distorsions, based on the experimental FIRAS bounds, and obtained by integrating equations~(\ref{eqs:mu_y}) as explained in section~\ref{sec:dist}. Finally, we report the BBN bounds obtained in Ref.~\cite{Poulin:2015opa}.

The complementarity of the CMB anisotropy constraints with those from CMB spectral distortions and BBN is obvious. The constraints from the angular power spectra dominate for long lifetimes, $\Gamma^{-1}\gtrsim 10^{12}\,$s. They reach a remarkable sensitivity around  $\Gamma^{-1}\sim t_c \equiv \Gamma_c^{-1}\equiv 10^{14}\,$s, excluding values of $\Xi$ as low as $\sim10^{-11}-10^{-10}$. For smaller lifetimes, they degrades roughly by a factor $\Gamma_c/\Gamma$.
Hence the CMB is maximally sensitive to particles decaying during the dark ages following photon decoupling, around a redhsift $z_c\simeq 300$ corresponding to the time-redshift conversion of $t_c$.

This suggests that the impact of e.m. decay on CMB anisotropies depends in first approximation on the amount of energy deposited around the time $t\sim t_c$, i.e. around the redshift $z_c$. A posteriori, we 
can thus propose $f(z) \longrightarrow f(z_c) e^{-\Gamma(t(z)-t_c)}$ as a physically motivated ansatz for the on-the-spot scheme:
his approximation captures the quasi-exponential shape of $f(z)$, and goes through the true value of $f(z)$ near the most relevant epoch $t \simeq t_c$. Following the definition of $f_{\rm eff}$ in Eq.~(\ref{eq:f_z_approx}), this corresponds to $f_{\rm eff}=f(300) e^{\Gamma t(300)}$.

Figure~\ref{Fig:AllConstraints} summarises the accuracy of this fast approximation scheme (dashed lines). The on-the-spot bounds are found to be biased at the $10-50\%$ level for the channels,  energies and lifetimes ($\gtrsim10^{14}$s) considered here. For shorter lifetimes, one would need to change Ansatz, which we did empirically in sec.~\ref{sec:CMBconstraints} by adjusting $f_{\rm eff}$ to $f(z_{\rm decay})e^{\Gamma t(z_{\rm decay})} = f(z_{\rm decay})e$ for $\tau\simeq 10^{13}\,$s. With this new Ansatz, we checked that the on-the-spot bound (not shown explicitly on the plot) is correct at the $20-50\%$ level. For lifetimes below $10^{13}\,$s, as illustrated in sec.~\ref{sec:CMBconstraints}, the on-the-spot approximation fails at describing correctly the physical effects of the decay. 
We conclude that the on-the-spot approximation is barely sufficient for deriving order-of-magnitude estimates of the bounds in the range $\tau \geq 10^{13}\,$s.

Not surprisingly, earlier universe probes are more sensitive to shorter lifetimes, although the bounds are never as strong as from the CMB angular power spectra for $\Gamma^{-1}\sim 10^{14}\,$s. It is remarkable that cosmological probes, combined, achieve a sensitivity over more than 20 orders of magnitude in lifetime! 
A few comments are in order:
\begin{itemize}
\item
There exist other bounds of astrophysical nature (e.g. from gamma-ray flux measurements, see for instance~\cite{Cirelli:2012ut}), which we do not discuss here in details. For energy injection happening mostly in the late universe ($z\ll 1$) and involving sufficiently high energy particles ($E\gg\,$MeV), these bounds can be more constraining than the cosmological ones. This is why we do not indulge here in the otherwise interesting parameter space in the upper-right part of Fig.~\ref{Fig:AllConstraints}.
\item 
Although we kept the point implicit until now, different probes are sensitive to different energies ranges. BBN is the most limited in this sense, since only photons or $e^\pm$ more energetic than the disintegration thresholds of light nuclei (from few to few tens of MeV) may have an effect. The CMB angular spectra constraints apply in principle to a much broader range of energies, virtually as low as $\sim 10.2\,$eV, below which no ionization can be induced. However,  the range of validity of the transfer functions can be questioned below the keV scale, and the factorization approximation of Eq.~(\ref{eq:factorization}) might not be reliable below the $\sim$10 MeV scale~\cite{Slatyer16}. In practice, unless one wants to run a ``case by case'' simulation, one is limited by the regime of validity of the approximations used. Finally, the CMB spectral distortions, although yielding the least restrictive of the constrains reported here, are in principle sensitive to a wide range of injected energies, even very low ones (although, to the best of our knowledge, bounds have always been derived under the restrictive assumption of injected particles being more energetic than the ones of the medium). Thus, even if we summarized the bounds in a larger parameter space with one extra dimension for the injected energy, the different cosmological probes would remain very complementary.
\end{itemize}

 Let us finish this section by a comparison with previous works. In Ref.~\cite{Slatyer12}, CMB anisotropies constraints have been derived for DM decaying into $e^\pm$ and photons, for several masses and lifetimes. We can compare our results to their Fig. 7, in which a forecast for Planck was presented. This forecast turns out to be in fairly good agreements with our results for lifetimes $\gtrsim 10^{13}\,$s, with a factor $\sim$2 difference in the long lifetime limit. However, differences are larger for very short lifetimes. We attribute them to i) the new energy deposition functions; ii) the known fact that Planck slightly underperformed with respect to earlier expectations regarding polarization errors. Nonetheless, Planck has improved by nearly one order of magnitude over the pre-existing WMAP 7 constraints. Another estimate of these constraints in the on-the-spot approximation has been made in Ref.~\cite{Zhang07}. If we take into account the limitations of their treatment, our results are in agreement with their Planck forecast (c.f. their Fig. 2).

\section{Applications and forecasts}\label{sec:Applications}
\subsection{Primordial black holes}\label{PBH}
A notable application of our results concerns the possibility of constraining the abundances of primordial black holes (PBH) at high ($\gtrsim~M_{\odot}$) and low ($\lesssim 10^{-15} M_{\odot}$) masses. In the high mass case, the emission of high energy electromagnetic particles in the plasma is typically due to the accretion of matter onto the PBH. This very complicated astrophysical process, studied for instance in some detail in \cite{Ricotti:2007au} is currently a hot topic, after the aLIGO discovery of two events binary BH mergers with masses in the 10 $M_{\odot}$ scale. 
A detailed revisitation of this regime is far beyond the scope of this paper and we postpone a reanalysis of these constraints to further work, limiting ourselves to note that the formalism and code development implemented here is one of the tasks needed towards that goal. On the other hand, the ``low mass'' PBH regime, for which energy injection is due to the evaporation of the PBH (the so-called Hawking radiation) is rather well understood and straightforward to implement. Quite interestingly, with the exceptions of Refs.~\cite{Belotsky:2014twa,Carr:2009jm} on which we will comment below, CMB constraints to this scenario have been overlooked in the past. A proper computation, to the best of our knowledge, was never performed so far, which is rather surprising since the bounds we derive (see below) turn to be competitive with the best ones in the mass range $10^{15}\,$g to $10^{16.6}\,$g,  which are given by the extragalactic gamma-ray background (EGB)~\cite{Carr:2009jm,Carr:2016drx}~\footnote{ Note that other gamma-ray bounds exist, like the Galactic ones considered in~\cite{Carr:2016hva}. While potentially stronger, they are more model-dependent, as also argued in~\cite{Carr:2016drx}, and we will not consider them in the following.}, and to be the dominant ones in the range $10^{13.3}\,$g to $10^{14.4}\,$g. 

\begin{figure}[!h]
\hspace{-0.8cm}
\includegraphics[scale=0.35]{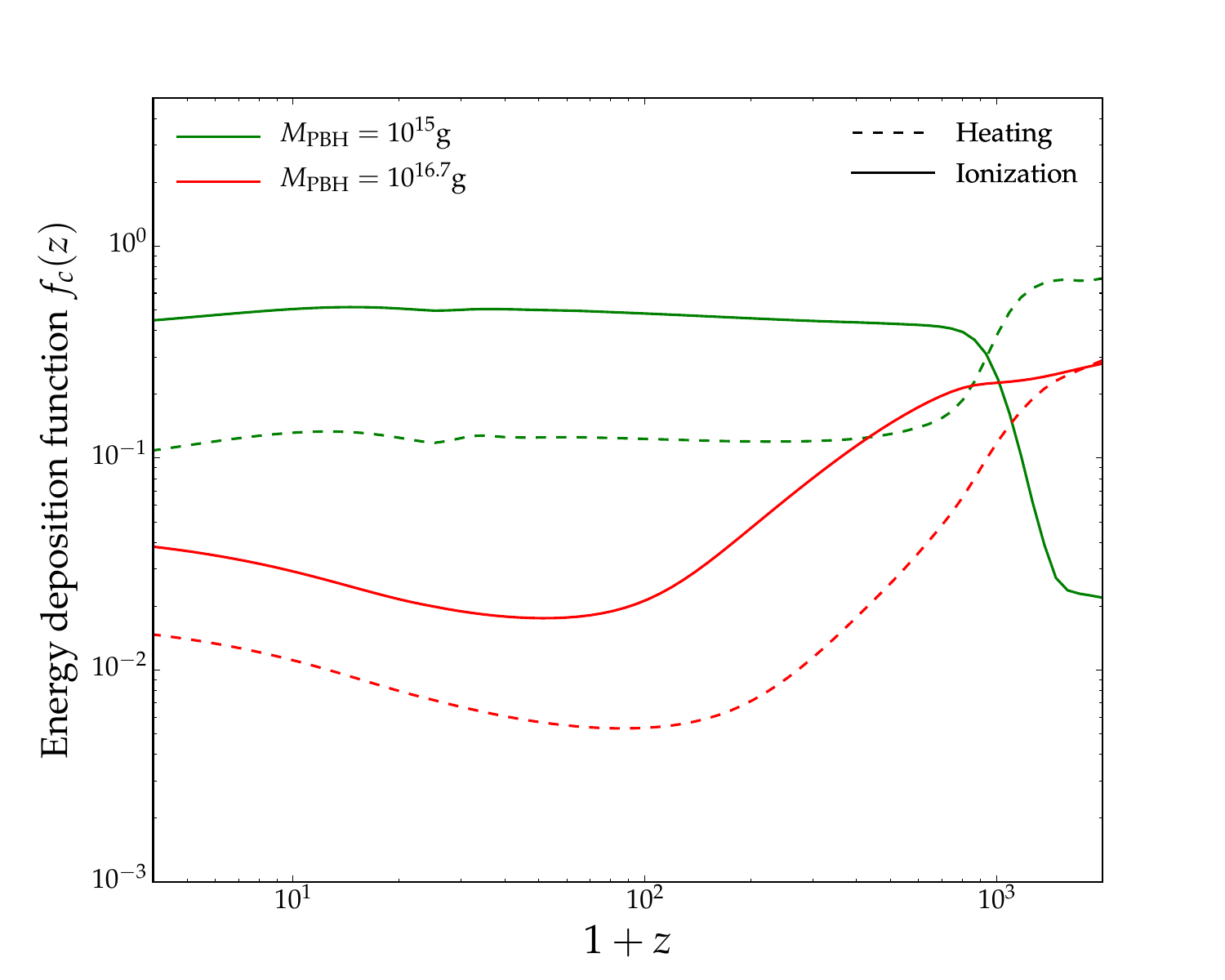}
\includegraphics[scale=0.35]{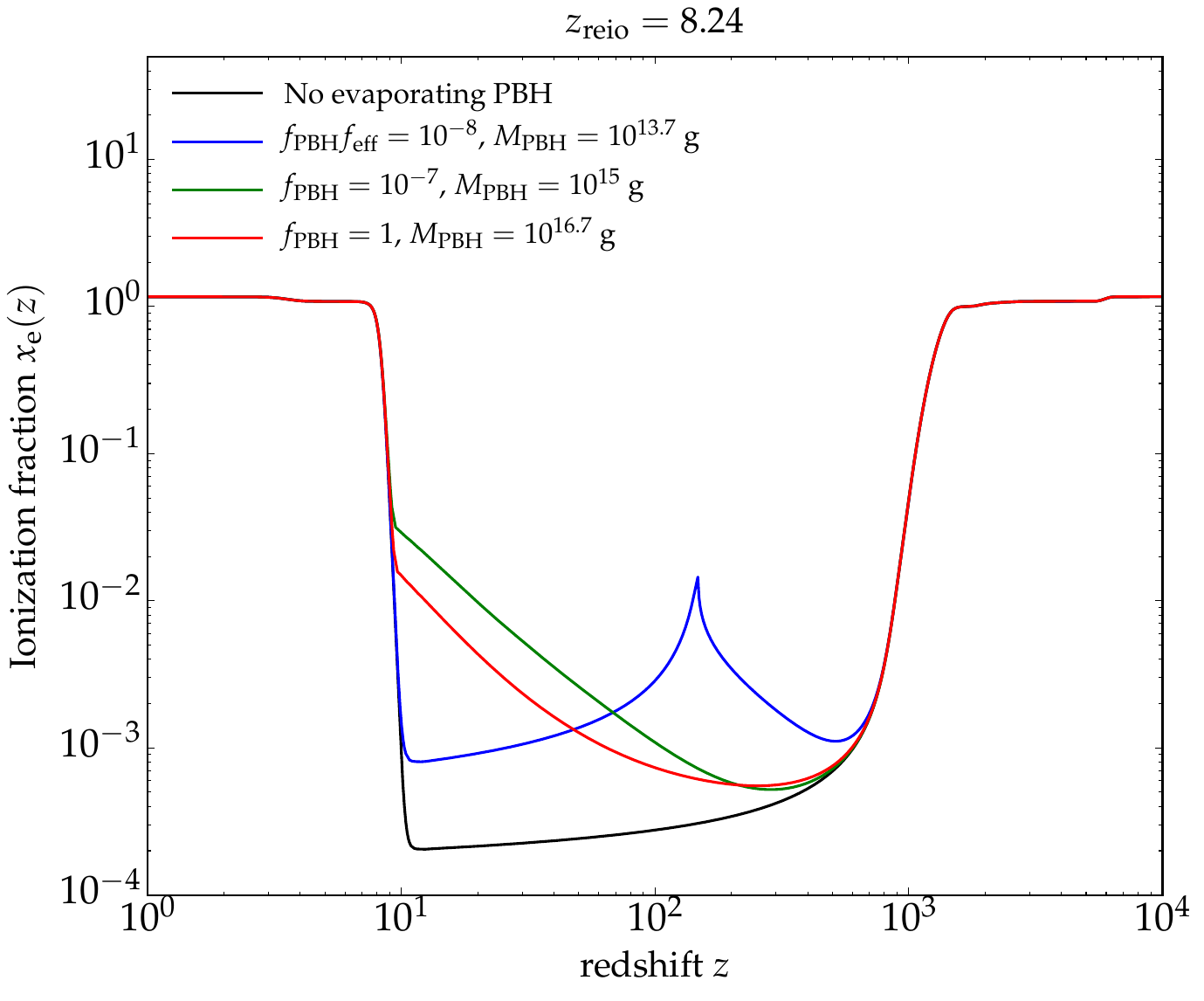}
\caption{\label{Fig:f_z_pbh}Left panel $-$ Energy deposition functions computed following ref.~\cite{Slatyer15-2} in the case of evaporating PBH of masses $1\times10^{15}$g and $10^{16.7}$g. Right panel $-$ Corresponding free electron fractions. }
\end{figure}

We summarize here the basic results of PBH evaporation, following ref.~\cite{Carr:2009jm}.
In a seminal couple of papers \cite{Hawking:1974rv,Hawking:1974sw}, Hawking showed that a Schwarzschild black hole of mass $M$ 
should emit black body spectra of particles at a temperature 
\begin{equation}
T_{\rm BH} = \frac{1}{8\pi G M}\simeq 1.06 \bigg(\frac{10^{13}\,{\rm g}}{M}\bigg){\rm GeV}\,.
\end{equation}
The mass-loss rate of an evaporating black holes can be expressed as 
\begin{equation}
\frac{dM}{dt}=-5.34\times10^{-11}{\cal F}(M) \bigg(\frac{M}{10^{13}\,{\rm g}}\bigg)^{-2}{\rm s}^{-1}\,,\label{mloss}
\end{equation}
where the function ${\cal F}(M)$, normalized to 1 if $M > 10^{17}\,$g, counts the number of (relativistic) particles species emitted by the PBH, and increases with decreasing mass. For instance, at $10^{15}\,$g also electrons and positrons (besides photons, neutrinos and gravitons) can be emitted, and ${\cal F}(M)\simeq 1.568$~\cite{MacGibbon:1991tj},
which further rises (due to muon production) to about 2.1 for $M\simeq 10^{14}\,$g. The number of degrees of freedom quickly rises for lower masses (when temperatures above the QCD phase transition one are attained) with ${\cal F}$ saturating at 15.35 when $M\lesssim {\rm few}\times 10^{10}\,$g, assuming the standard model to be the correct theory.
By integration of Eq.~(\ref{mloss}) over time, one obtains the lifetime
\begin{equation}\label{eq:Gamma}
\Gamma_{\rm PBH}^{-1}\simeq 4.07 \times 10^{11}\bigg(\frac{{\cal F}(M)}{15.35}\bigg)^{-1}\bigg(\frac{M}{10^{13}{\rm g}}\bigg)^{3}{\rm s}\,.
\end{equation}
and the energy injection rate for low mass PBH is given in full generality by 
\begin{equation}
\frac{dE}{dVdt}\bigg|_{\rm inj,~PBH}=\frac{\Omega_{\rm DM}\rho_cc^2(1+z)^3f_{\rm PBH}c^2}{M_{\rm PBH}^{\rm ini}}\frac{dM}{dt}\bigg|_{\rm e.m.}~.
\end{equation}
where $dM/dt\big|_{\rm e.m.} = f_{\rm e.m.}dM/dt$ with $f_{\rm e.m.}$ the e.m. branching ratio of an evaporating PBH, which depends on its mass.
 If $M\gtrsim 10^{15}\,$g, the PBH lifetime is longer than the age of the universe, and its mass and lifetime are roughly time-independent parameters: for instance, a PBH of initial mass $10^{15}$ g would weight $9.6\times10^{14}$ g today. 
Since we do not aim anyway at better than ${\cal O}$(10$\%)$ level accuracy, in this mass range we assume that both $M$ and the lifetime $\Gamma^{-1}$ are time independent, which leads to a simple exponential decay law, similar to the case of decaying particles with lifetime given by Eq.~(\ref{eq:Gamma}).
On the other hand, in this range, we do calculate the ionization and heating efficiency on the basis of a reasonable approximation for the spectrum of emitted particles.
The spectra of particles emitted by the BH per unit of time with energy between $E$ and $E+dE$ are given by:
\begin{equation}
\frac{d\dot{N_s}}{dE}\propto\frac{\Gamma_s}{e^{E/T_{\rm BH}}-1(-1)^{2s}}\,,\label{emittPBHpart}
\end{equation}
where $s$ is the spin of the particle, and the dimensionless absorption coefficient $\Gamma_s$ can be written  in the high-energy limit $E\gg T_{\rm BH}$ as \cite{Page:1976df}:
\begin{equation}
\Gamma_s(M,E)=27E^2G^2M^2\,.
\end{equation}
It is also known that the spectra fall rapidly at low energy.  To avoid unnecessary complication, in what follows we thus approximate the spectra as vanishing below $E=3\,T_{\rm BH}$, while  sticking to the high energy limit formula above this energy. Note that a slightly different choice would not have changed our conclusions significantly, since in most cases the energy deposition functions are not strongly dependent on the energy. Also note that we do not need to worry about the normalization in Eq.~(\ref{emittPBHpart}), since we use it to compute the deposition functions, which only depend on the energy shape (See Eq. (2.8)). Put otherwise, the correct normalization is assured by using the appropriate $M$, $\Gamma$, and the correct branching ratio in e.m. channels.
 On the other hand, the above-mentioned approximations for the spectra assume relativistic particles. For $e^{\pm}$, this becomes too crude for masses above $\sim 10^{16.8}\,$g, hence we limit our considerations to lighter PBHs. Also note that EGB constraints tend to dominate anyway at these masses, and beyond that GRB femtolensing constraints at the level of $\Omega_{\rm PBH}\lesssim 0.1\,\Omega_{\rm DM}$ take over~\cite{2012PhRvD86d3001B,Carr:2016drx}, making a detailed computation of CMB bounds less appealing for such high masses.  For a few cases in the mass interval $10^{15}\,{\rm g} \leq M\leq 10^{16.8}\,{\rm g}$ we recompute the energy deposition functions $f_c(z,x_e)$
without relying on the factorization hypothesis, since it breaks down for some of the injection energies of interest. We conservatively assume the standard reionization scenario to
compute the $f_c(z,x_e)$'s~\footnote{This is not strictly correct, but only approximately true as long as $x_e$ stays very small. Also, an iterative treatment, starting from a \texttt{recfast}-like ionization history and recomputing the transfer functions once the effects of the decay on the ionization fraction have been included, showed not only a quick convergence but that the approximate results are, if anything, conservative~\cite{Liu:2016cnk}.}. 
Our results for $M = 10^{15}$g and $10^{16.7}$g are shown in Fig.~\ref{Fig:f_z_pbh}, together with the resulting free electron fractions.

For masses $M\lesssim 10^{13}\,{\rm g}$, the evaporation takes place before $10^{12}\,$s, hence we expect CMB anisotropies to be insensitive in that range (see Fig.~\ref{Fig:AllConstraints}). In the range $ 10^{13}\,{\rm g}\ll M\ll 10^{15}\,{\rm g}$, the actual time dependence of the evaporation is important. The emitted spectrum is rapidly changing with the mass, which complicates substantially the treatment, in particular in terms of computational time. 
 While leaving a complete computation of the constraints to a future work, it is however interesting to show how the non-trivial time dependence of the mass affects the bounds. To make this effect apparent, we simplify the treatment of the energy deposition function and resort to the  on-the-spot approximation. Indeed, for PBHs with masses $\gtrsim 10^{14.5}\,$g, most of the evaporation happens  at times $\gtrsim 10^{14}\,$s. Hence, we expect the effective criterion derived in sec.~\ref{sec:OtherConstraints} to apply relatively well. We can estimate the efficiency of this approximation by comparing with the full treatment for masses $>10^{15}\,$g, finding a reasonably good agreement. 

In order to compute a value of the effective energy deposition efficiency in the on-the-spot limit, we perform further approximations: a) we only consider the branching ratio into  $e^\pm$ and $\gamma$, neglecting the energy eventually deposited by the decay products of heavier particles like muons and pions (if kinematically allowed). b) We fix this branching ratio to the {\em initial} one (i.e. given the initial mass of the PBH).  c) We neglect the evolution in energy of the injected spectra, fixing it to the initial value. Eventually, in the very last stages of the life of the PBH all standard model particles are produced, and the energy spectra of the emitted particles change, so all these approximations break down. However, most of the PBH mass has been radiated away at an earlier epoch, hence on a purely ``calorimetric'' ground our choices are reasonable and should capture the bulk of the effect.  Note that approximation a) should slightly underestimate the bounds. Approximation b) should push towards overestimating a bit the bound, while approximation c) can play in either sense, but is expected to be rather mild: For most of the time the typical particle energies range from $\sim $ 10 MeV to few hundreds of MeV, hence the time evolution in deposition functions should not be extreme. By performing such simplifications, one can readily integrate Eq.~(\ref{mloss}) and find an analytical solution (different from previous constant decay rate) for the time evolution of the PBH mass. We plot the free-electron fraction for the illustrating case $M_{\rm PBH}=10^{13.7}\,$g in Fig.~\ref{Fig:f_z_pbh}--left panel. It shows a very singular
 behavior with a peak coming from the peculiar energy injection history, making it in principle distinguishable from an exotic particle decay.

\begin{figure}[t]
\centering
\includegraphics[scale=0.5]{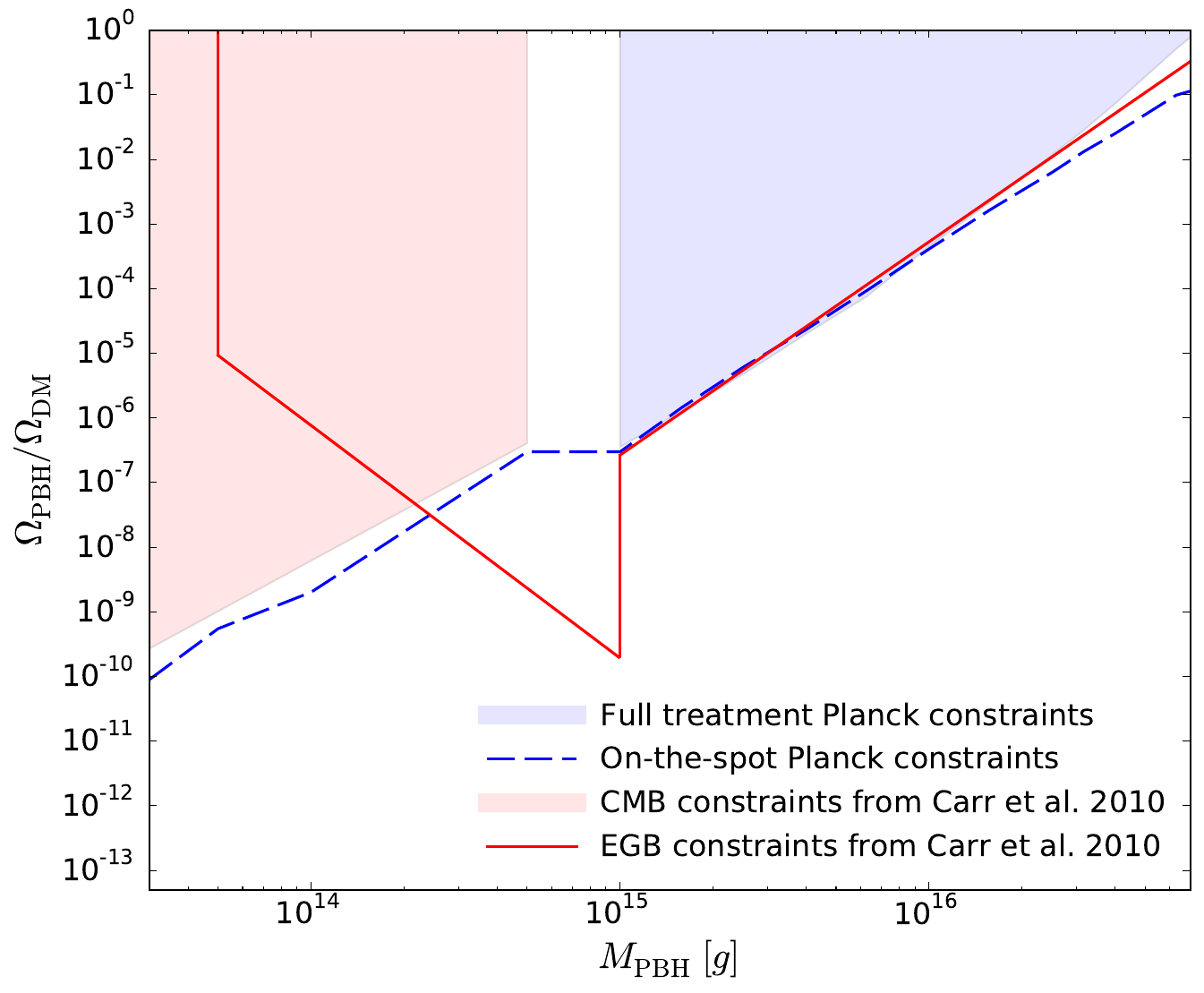}
\caption{\label{Fig:PBH_Constraints}Constraints on the abundance of PBH normalized to the DM one as a function of the PBH mass with the full treatment (shaded blue area) and its approximate version (blue, dashed). Limits from extragalactic $\gamma$-ray background (solid, red) and CMB anisotropies (pink shaded area) from Ref.~\cite{Carr:2009jm} are also shown.}
\end{figure}

Using {\sf CLASS} and {\sf MontePython}, we run $\Lambda$CDM + $f_{\rm PBH}\equiv\Omega_{\rm PBH}/\Omega_{\rm DM}$ for 10 different PBH masses distributed between $10 ^{13.5}\,$ g and $10 ^{16.8}\,$g, using Planck high-$\ell$ TT,TE,EE + simlow (prior on $\tau_{\rm reio}$) and lensing likelihood.  The results of the MCMC scan in the plane $\{f_{\rm PBH},M_{\rm PBH}\}$ are shown in Fig.~\ref{Fig:PBH_Constraints}, together with constraints coming from the EGB from Ref.~\cite{Carr:2009jm}, as well as CMB constraints from Ref.~\cite{Zhang07} applied to the case of low-mass PBH in Ref.~\cite{Carr:2009jm} via a simple prescription. Our constraints turn out to be very competitive with $\gamma$-ray background one in the range $10^{15}\,$g to $10^{16.6}\,$g and to dominate in the range $10^{13.5}\,$g to $10^{14.4}\,$g. 
We do not extend the study to lower masses since the on-the-spot approximation is  known to fail. We expect however the constraints to rapidly degrade at lower masses.
Note that the constraint in the low mass range, while very strong, is not simply the prolongation of the high-mass one: the ``shoulder'' below $10^{15}\,$g is due to the combined effect of new channels like muon pairs opening up (which, being less effective in releasing energy, lower the ``useful'' e.m. branching ratio), and on the slightly less efficient energy deposition at the correspondingly higher injection energies. Also note that the similarity of the constraints with those derived in Ref.~\cite{Carr:2009jm}  is accidental: the data available almost a decade ago where significantly less constraining, but the treatment in~\cite{Carr:2009jm} overestimated the constraining power due to a number of approximations: for instance they did not follow the proper time-evolution of the mass; they did not estimate the efficiency of the energy deposition (they implicitly worked with $f_{\rm eff} = 1$) which overestimates the energy deposition by a factor 2 to 3 depending on the PBH mass. Finally, our constraints are not nearly as good as $\gamma$-ray background one in the range $10^{14.4}\,{\rm g}-10^{15}\,$g. That said, there is still room for improvement with respect to our current treatment, notably for masses below $10^{15}\,$g.

Let us finish this section with a quick comparison with Ref.~\cite{Belotsky:2014twa}. In this work, an estimate of the impact of PBH within the mass range $10^{16}\,{\rm g}-10^{17}$ g was made. They assumed an effective ``on-the-spot'' approximation, working out the value of the absorption efficiency (which would roughly correspond to our $f_{\rm eff}$) under some simplified assumptions for the energy losses and spectra of evaporated particles. However, what might lead to the biggest difference with our work is that they compute $x_e(z)$  using Saha formula. Hence, the energy injected to the medium affect the ionization history only through reheating, which in turns lead to very different evolution for $x_e$ as can be seen from our Fig.~\ref{Fig:f_z_pbh}$-$ right panel, compared to their Fig.~6. 
Our more accurate computation, that takes into account both the impact of energy injection through heating and the  ionization (as well as excitation) of the atoms, shows that the free electron fraction evolution departs significantly from Saha equation, with a reionization starting already at redshift of few hundreds. Thus, the constraints of Ref.~\cite{Belotsky:2014twa}
should not be considered quantitatively reliable.

\subsection{Sterile neutrinos}
\label{sec:neutrinos}
\begin{figure}[t]
\centering
\includegraphics[scale=0.5]{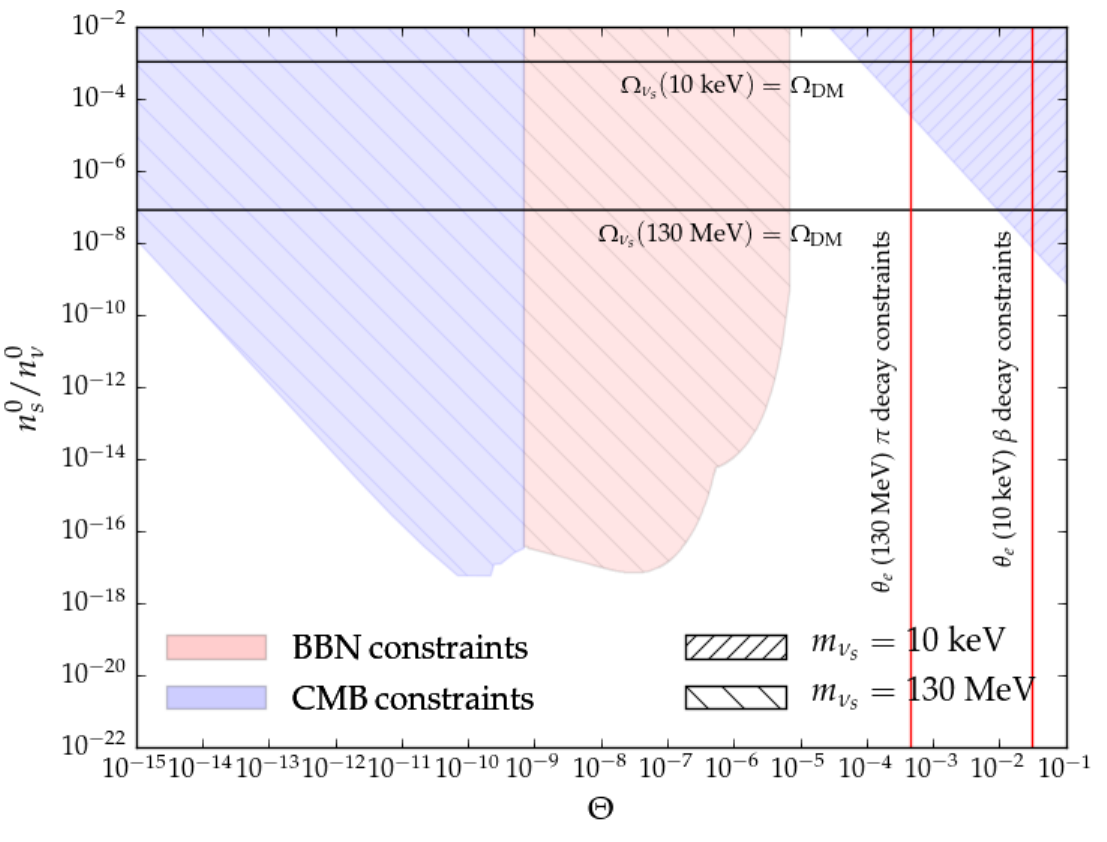}
\caption{\label{Fig:sterile_neutrinos}Cosmological constraints on the abundance of sterile neutrinos (including antineutrinos) normalized to the total active neutrino density, with no prior on their production mechanism. We consider a sterile neutrino mass of $10\,$keV, the typical scale for sterile neutrinos being DM candidates, as well as the heavier scale of $130$ MeV
for much weaker couplings, a parameter space usually ignored.The solid horizontal lines indicate the condition $\Omega_{\nu_s}=\Omega_{\rm DM}$, which is only a function of the mass: The bounds above those lines, for the corresponding shaded areas, are only nominal, since the parameter space is most excluded by the overclosure condition $\Omega_{\nu_s}>\Omega_{\rm DM}$. Red vertical lines correspond to direct laboratory constraints from nuclear $\beta$ decay \cite{deGouvea:2015euy} and $\pi$ decay \cite{Britton:1992xv}, assuming mixing with electron neutrino species.}
\end{figure}

The discovery that neutrinos oscillate, hence that at least two of them are massive, is one of the few hints (and the only laboratory one) for physics beyond the standard model.
The most minimal extension of the SM is probably the one requiring right-handed neutrinos, which are SM gauge singlets (hence `sterile') and so weakly interacting that usually astrophysical and cosmological observables are better suited to probe them than laboratory ones. 
The possibility that they may be related to other mysteries like the baryon asymmetry of the universe or the DM one is well discussed in the literature (see e.g.~\cite{Adhikari:2016bei} for a recent review). If these neutrinos are of Majorana type, they can have an extra mass term whose parametric size is a priori unconstrained by theoretical considerations. 
Obviously, the weaker their coupling, the harder they are to constrain. These neutrinos are however unstable: provided that they light enough, cosmology can help, since it can probe very long timescales.  As long as the sterile neutrino ($\nu_s$) mass is below $\sim 1\,$MeV, the only channels open are
\cite{Shrock:1980ct,Pal:1981rm,Johnson:1997cj,Gorbunov:2007ak}
\begin{itemize}
\item[\textbullet] $\nu_s\to3\nu$, with $\Gamma_{3\nu}\simeq\frac{G_F^2M_s^5\Theta^2}{192\pi^3}\simeq \left[3\times 10^{4}\,{\rm s}\left(\frac{\rm MeV}{M_s}\right)^5\Theta^{-2}\right]^{-1}$;
\item[\textbullet]$\nu_s\to\nu_\alpha+\gamma$, with $\Gamma_{\nu\gamma}\simeq\frac{9G_F^2\alpha M_s^5\Theta^2}{256\pi^4}\simeq 0.016\,\Gamma_{3\nu}$.
\end{itemize}
In the above formulae, we introduced $\Theta^2 \equiv \sum_{\alpha}\theta_\alpha^2$, where $\theta_\alpha$ is the mixing angle with active neutrino of flavour $\alpha$. When the $\nu_s$  mass exceeds $2\,m_e$ a third channel, $\nu_s\to \nu_\alpha e^+e^-$, is also open, with a rate depending on single $\theta_\alpha$'s, but in general with a b.r. of $\sim 30\%$ as long as its mass does not exceed $m_{\pi^0}\simeq 135$ MeV above which pionic channels are also open. These relations imply that, as long as $\left( M_s/{\rm MeV}\right)^5\Theta^{2}\ll 1$, these particles decay at ``cosmologically interesting times'', with an associated b.r. into e.m. channels ranging  from 1.6\% at masses below the MeV to {\cal O}(30\%) at masses up to $\sim 130$ MeV.
The CMB is capable of constraining such particles even if they only constitute a negligible fraction of the DM. It is straightforward to translate the $\Xi$-lifetime constraints derived in Fig. \ref{Fig:AllConstraints} in more meaningful parameters for sterile neutrinos, such as their mixing angle and relative abundance with respect to the total active neutrino species (including antineutrinos), which we do in Fig.~\ref{Fig:sterile_neutrinos}  for two values of the masses. Note that we have assumed monochromatic spectra for the daughter particles. This is basically exact in the 10 keV case, for which the only visible decay mode is $\nu_s\to\nu_\alpha+\gamma$ and the doppler broadening is negligible. In the 130 MeV case, the dominant e.m. decay mode is  $\nu_s\to\nu_\alpha e^+e^-$, which was treated as described in Sec.~\ref{sec:standard_eq}. For the typical 10 keV mass scale required for sterile neutrino to be the DM, we obtain bounds comparable to \cite{Oldengott:2016yjc}, which are typically weaker than bounds from astrophysical probes, notably  X-ray lines.  In general, however, even higher masses and lower abundances can be constrained. For instance, a 130 MeV sterile neutrino would represent a sizable amount of DM only if its population is larger than a fraction $\gtrsim 10^{-8}$ of the active neutrino one~\footnote{Note that in the upper part of the parameter space of Fig.~\ref{Fig:sterile_neutrinos}, above the solid horizontal lines indicating the condition $\Omega_{\nu_s}=\Omega_{\rm DM}$ (which is only a function of the mass), our constraints are only nominal, since we are assuming no major departure from the standard history of $\Lambda$CDM. That parameter range is likely excluded by an unacceptably large modification of the expansion history of the universe due to the ``overclosure'' condition $\Omega_{\nu_s}>\Omega_{\rm DM}$, although we do not compute these constraints in detail here.}, yet the CMB can constrain a fraction up to  a billion times smaller than that, for incredibly tiny mixing angles smaller than $10^{-10}$! To put the latter constraints into a context, one may compare them with the ones reported in Fig. 7 of Ref.~\cite{Abazajian:2001nj}. In particular, our new constraints are relevant in the upper-left corner of the parameter space (labelled as $\tau<t_{\rm today}$)  i.e. the region where sterile neutrino relics from the early universe were indicated as decaying ``without constrainable effects and make no contribution to the present matter density''. 
Note that the laboratory constraints from $\beta$ decay nuclei \cite{deGouvea:2015euy} and $\pi$ decay \cite{Britton:1992xv}, are far from being competitive with our cosmological ones.  For reference, in Fig.~\ref{Fig:sterile_neutrinos} we have indicated with vertical red lines the best laboratory upper bounds available for the two masses assuming mixing with electron neutrino species, since direct bounds on other mixings are much weaker: For the heavier mass case, see \cite{Yamazaki:1984sj,Orloff:2002de};  for the 10 keV neutrino mass case, it is argued in \cite{deGouvea:2015euy} that no model-independent laboratory bounds exist on those mixings.

Are these parameter values of any interest, given the masses and mixing knowledge on active neutrinos? In fact,  some relations between mass-scale and mixing angles exist, once the phenomenological knowledge on the smallness of (active) neutrino mass is factored in. This is nicely reviewed in~\cite{Alekhin:2015byh}, notably in its chapter 4. The Yukawa couplings, or equivalently a combination the mixing angle elements $U$ and masses $M_I$ of the right-handed neutrinos should obey the relation
\begin{equation}
\sum_{\alpha,I} U_{\rm \alpha I}^2\,M_I^2\geq M_{\rm min}\,\sum m_\nu=10^{-5}{\rm MeV}^2 \frac{m_\nu}{0.1\,{\rm eV}}\frac{ M_{\rm min}}{1\,{\rm MeV}}\,,
\end{equation}
where $M_{\rm min}$ is the lightest sterile neutrino contributing to the mass of the standard model neutrinos.
Ignoring for the moment any cosmological or particle physics constraints, this relation together with the empirically constrained mass scale of neutrinos suggests that
the lightest extra neutrino mass eigenstate involved in the active neutrino mass mechanism should have an effective mixing element  $U_{\rm \alpha I}^2\simeq \Theta^2\gtrsim 10^{-5}M_{\rm MeV}^{-1}$. 
We conclude that cosmology is typically sensitive to neutrinos more weakly coupled than those implied in the mechanism giving mass to the active neutrinos.
This is a qualitatively important point: while the existence of {\it two} right handed neutrinos is needed to account for the observed oscillation phenomenology, and 
is remarkably sufficient to account for leptogenesis via the oscillation mechanism first described in~\cite{Akhmedov:1998qx},  the number, masses and mixing of additional right-handed neutrinos are not
otherwise constrained by particle physics constraints or fundamental physics conditions, rather must be suggested by observational arguments.
For instance,  in the so-called $\nu$MSM~\cite{Asaka:2005an,Asaka:2005pn},
a three right handed neutrino scenario is invoked, with the lightest one being responsible for the DM of the universe, not certainly for the particle physics consistency; the third state must then essentially be decoupled from the active ones, contributing negligibly to their mass generation mechanism.  This example shows that additional decoupled states might be only be relevant for cosmological consequences. Our calculations above shows that the potentially interesting parameter range to alter recombination or reionization history is much wider than the one invoked to account for sterile neutrino DM, extending to MeV or GeV masses.
Needless to say, from the conceptual point of view, having as many handles as possible on the existence of extra sterile neutrino states is important, since they could give us hints or constraints on the underlying symmetries of these extensions of the standard model.

\subsection{The 21 cm signal from the Dark Ages}
We have considered scenarios where e.m. energy injection takes place at some time in the history of the Universe. We saw that the CMB is particularly sensitive to
energy injected around or just after recombination, while BBN and CMB spectral distortions probe earlier times. What about cosmological probes of the more recent universe? 
Just after recombination, the universe is dark, cold and almost structureless, hence the name of ``Dark Ages'' for this period.
When $z$ drops below $\sim$300, however,  the baryons thermally decouple from the CMB photons, and start cooling more rapidly. The gas eventually warms up significantly above the CMB temperature around the epoch of reionization ($z\sim 10$), where a significant contribution of stellar and astrophysical sources is expected to take over the dynamical evolution of the gas. Thus, at redshift $z\lesssim$300 the conditions of temperature and ionization of the (mostly neutral) cosmic gas can be in principle probed via the hyperfine transition in  neutral hydrogen atoms:  it is natural to ask oneself whether significant exotic traces can be found in this highly redshifted 21 cm signal. 
Although the detection of the 21 cm signal from very high redshifts will probably stay beyond reach for a long time to come, prospects for the SKA experiment should allow detection capability up to $z\simeq 27$~\cite{Koopmans:2015sua}.  Such a sensitivity would be sufficient to open a yet unexplored window in the history of the universe. But is it useful to probe non-standard processes as well? For the most widely studied signals of DM annihilation, earlier results were rather encouraging about the discovery perspectives~\cite{Furlanetto:2006wp,Valdes12,Evoli14}. The authors of the recent paper~\cite{Lopez-Honorez:2016sur} have reinvestigated these forecasts with the most up-to-date tools, unfortunately finding them to be very challenging. In particular, due to the large  uncertainties in the reionization modeling, it appears hard to unambiguously isolate an exotic DM annihilation signal. 

\begin{figure}
\centering
\includegraphics[scale=0.5]{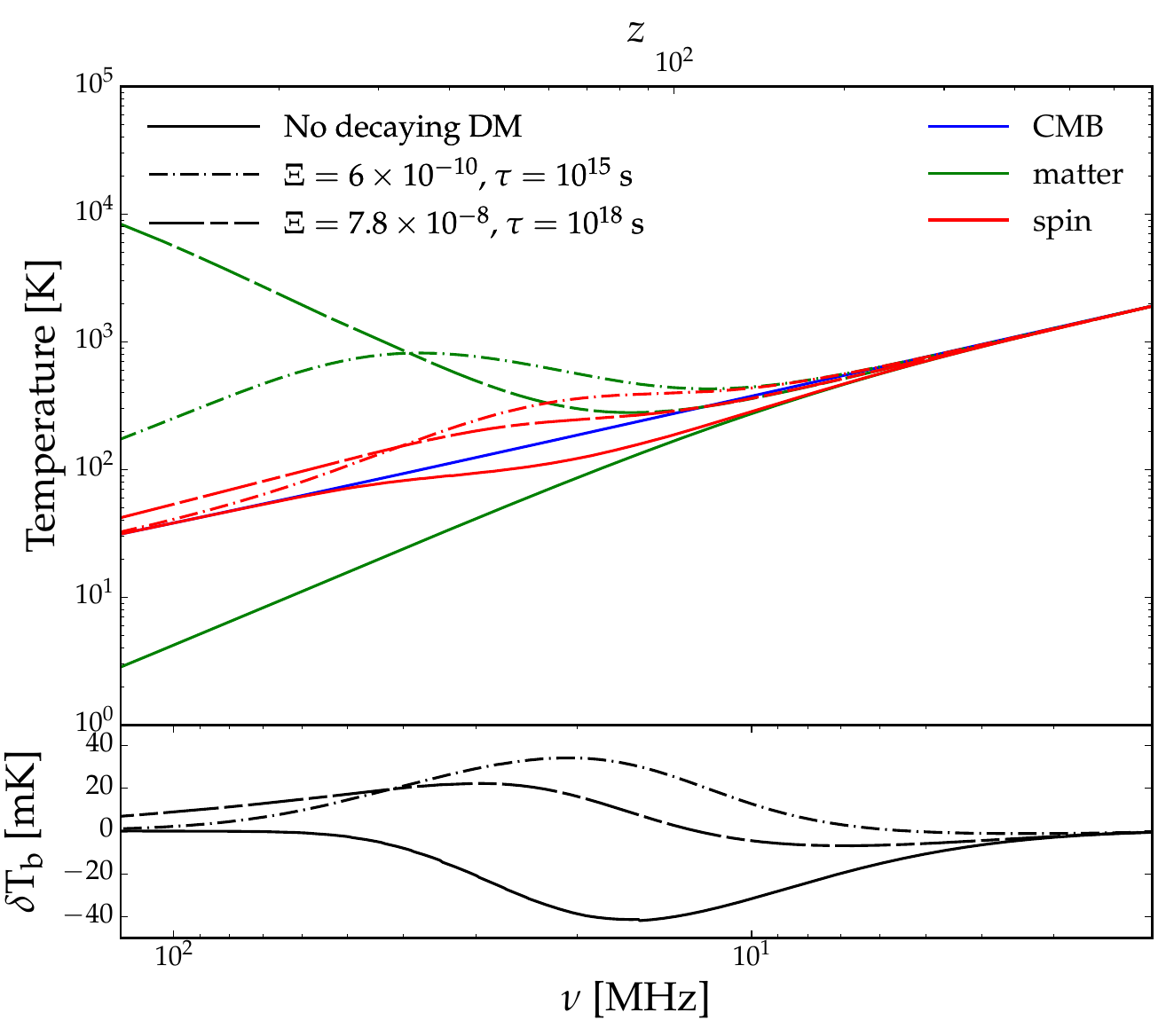}
\caption{\label{Fig:21cmSignal}Temperature history and mean differential brightness temperature during the Dark Ages for $\chi\to e^+e^-$ with decaying particle mass $m =200$ MeV, compared to the standard $\Lambda$CDM model. The decaying particle abundance is just within the currently allowed range.}
\end{figure}

\begin{figure}
\centering
\includegraphics[scale=0.5]{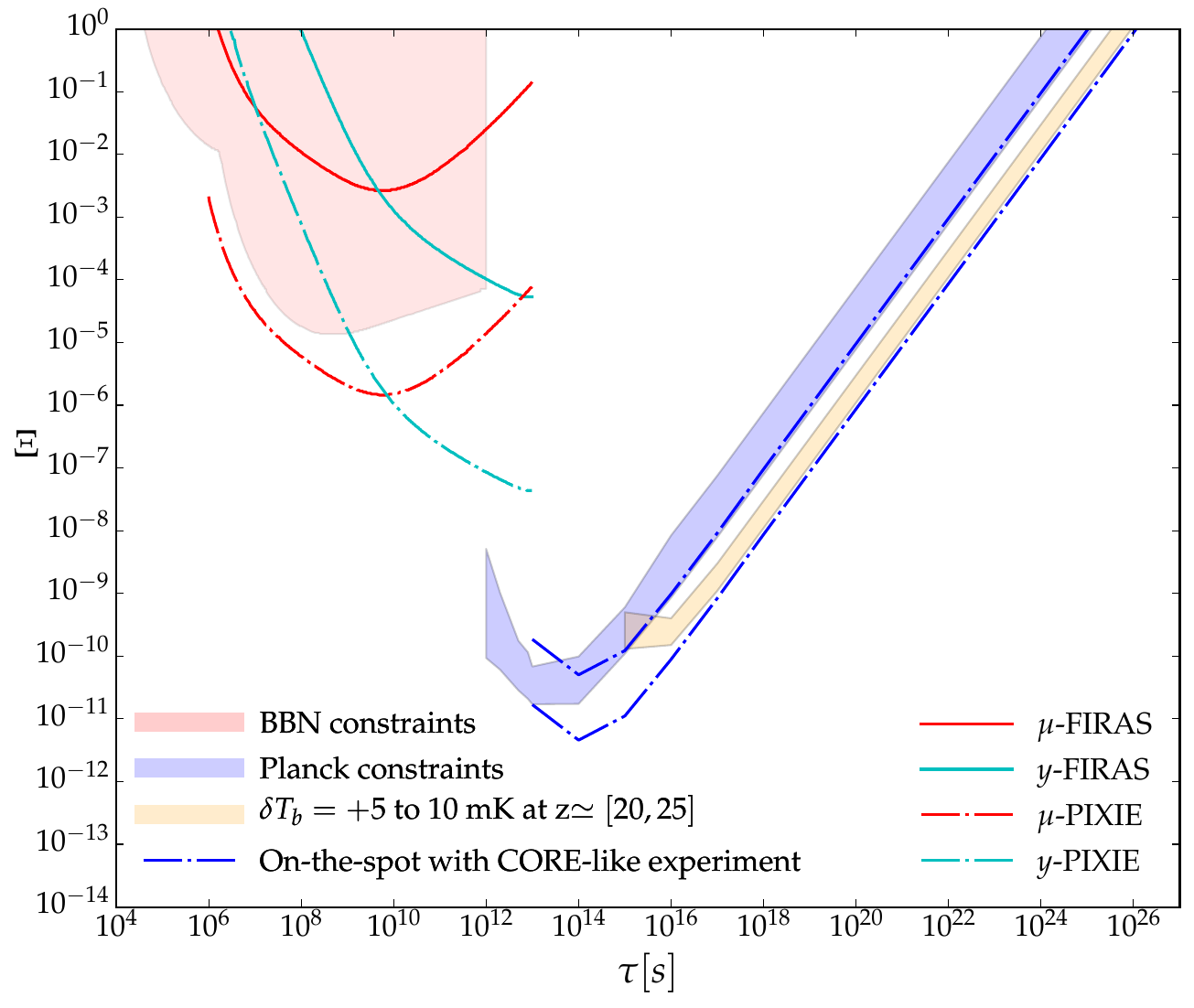}
\caption{\label{Fig:Forecast21cmSignal}A comparaison between current constraints on e.m. decaying exotic particles and a forecast using the sensitivity of SKA on the global differential brightness temperature $\delta T_b$ in the redshift range $[20,25]$, {\sf PiXiE} sensitivity on $\mu$ and $y$ distortions, as well as a CORE-like experiment. The blue shaded area, labeled as Planck constraints, now regroups $e^\pm$ and $\gamma$ results. The orange band is obtained for $\chi\to e^+e^-$ with decaying particle mass $m =200$ MeV.}
\end{figure}

However, the situation may be significantly better for energy injection processes concentrated at an earlier epoch, as for instance associated to a fraction of DM decaying in the so-called ``Cosmic Dawn'' period $15\lesssim z\lesssim 30$. The astrophysical processes are expected to be relatively mild at that epoch. In fact, observationally we know that they should not perturb too much the medium, e.g. triggering a too early ionization epoch, in order to be in agreement with the optical depth measurement by Planck. In addition, all modern parametric studies of astrophysical effects at this epoch indicate that they should be unable to reheat the gas above the CMB temperature, i.e. the 21 cm should be seen ``in absorption'' with respect to the CMB (see for instance~\cite{Mirocha:2015jra}, Fig. 1).  This is also the case for exotic signals such as annihilating DM in halos~\cite{Lopez-Honorez:2016sur} above redshift $z\simeq 20$. 
Here, we wish to briefly assess the possibility that models where a fraction of DM decays via e.m. channels, {\it not yet excluded by CMB} or other probes, can be uniquely tested via 21 cm observations at the Cosmic Dawn. While we certainly expect peculiar signatures in the power-spectrum (and possibly higher order statistics) of the 21 cm signal, for this preliminary study we will content ourselves with showing that a smoking gun signal is potentially present already at the level of the average differential brightness temperature $\delta T_b(\nu)$. This quantity is obtained by comparing lines of sight through a neutral hydrogen cloud to patches of the sky with clear view of the CMB. Following a textbook calculation (see e.g. Ref.~\cite{Furlanetto:2006jb}), one can easily compute the theoretical average signal (neglecting perturbations):
\begin{equation}
\delta T_b(\nu)  =\frac{T_S-T_{\rm CMB}}{1+z}\big(1-e^{-\tau_{\nu21}}\big) \simeq  27 x_{\rm HI}\bigg(1-\frac{T_{\rm CMB}}{T_S}\bigg)\sqrt{\bigg(\frac{1+z}{10}\bigg)\bigg(\frac{0.15}{\Omega_mh^2}\bigg)}\bigg(\frac{\Omega_bh^2}{0.023}\bigg)\,,
\end{equation}
where $x_{\rm HI}$ is the neutral hydrogen fraction and $T_S$ the spin temperature, the excitation temperature of the 21 cm transition. It is defined via $n_1/n_0=3 \exp(-T_{21}/T_S)$ with $T_{21} = 0.068\,$K. Assuming equilibrium between excitation processes (typically CMB photons absorption, collisions within atoms and scattering of UV photons from stars) and de-excitations ones, one can write a solution to the radiative transfer equation and get the evolution of the spin temperature:
\begin{equation}
T_S^{-1} = \frac{T_{\rm CMB}^{-1}+x_cT_{\rm M}^{-1}+x_\alpha T_c^{-1}}{1+x_c+x_\alpha}
\end{equation}
where $x_c$ and $x_\alpha$ are coupling coefficients for collisions and UV scattering respectively and $T_c$ is the effective color temperature of the UV radiation field. Radiative transfer typically drives $T_c$ to $T_{\rm M}$~\cite{Furlanetto:2006jb}, but in our case ($z>20$) we expect a negligible stellar contribution and simply set $x_\alpha$ to zero (this is a conservative assumption given the point we want to make).The collisional coupling $x_c$ can be computed as:
\begin{equation}
x_c=A_{10}^{-1}\frac{T_{21}}{T_{\rm CMB}}(n_H\kappa^{HH}_{10}+n_{e}\kappa^{eH}_{10})\,.\label{xcformula}
\end{equation}
In Eq.~(\ref{xcformula}), $A_{10}=2.85\times10^{-15}$s$^{-1}$ is the Einstein spontaneous emission coefficient of the 21 cm transition and the $\kappa^{i\,H}_{10}$ are the de-excitiation rates in hydrogen atom collisions~\footnote{The role of Helium is not expected to change qualitatively this picture and is neglected here.} with species $i$. They are tabulated in Ref.~\cite{Furlanetto:2006jb}.
We have computed the temperature history and the mean differential brightness temperature for decays of exotic particles with different lifetimes. The typical result is shown in Fig.~\ref{Fig:21cmSignal} for the two cases $\Gamma^{-1}=10^{15}\,$s and $\Gamma^{-1}=10^{18}\,$s. For definitiveness,  we consider a 2-body decay into electrons with the fraction of decaying DM fixed to its upper limit from previous analysis. With respect to the conventional evolution in the Cosmic Dawn, characterized by a negative $\delta T_b$ (see solid black curve in the bottom panel),
the models considered here show a {\it positive} $\delta T_b$, i.e. would be associated to a 21 cm signal seen {\it in emission} already at $z\simeq 20-25$ (for the case of PBH, this peculiar feature was already noted in~\cite{Mack:2008nv}). The effect is particularly noticeable for the shorter lifetime case, $\Gamma^{-1}=10^{15}\,$s, but remains appreciable also for timescales comparable with the universe lifetime, $\Gamma^{-1}=10^{18}\,$s, due to the small fraction of decays happening early on. The appearance of a signal in emission should constitute a smoking gun: if SKA were to observe an absorption signal from Cosmic Dawn, as expected, it would put constraints on  these exotic scenarios. At the same time, it also means that 21 cm studies have significant room for discovery. To gauge the level of the effect, it is worth bearing in mind that the SKA should have a sensitivity to $\delta T_b$ at ${\cal O}$(1-10) mK level at $z\simeq 25$, see Fig. 4 in~\cite{Koopmans:2015sua}. The orange band of Fig.~\ref{Fig:Forecast21cmSignal} shows the parameter range creating a 5 to 10 mK increase 
in $\delta T_b$ with respect to standard model expectation, which as argued would also lead to a change of sign in the signal. Here we limit ourselves to the specific case of the 2-body decay into electrons with decaying particle mass $m = 200$ MeV, but it is clear that even these crude considerations show great potential to go beyond the parameter space currently constrained.  Also note that we have not included informations on the power spectrum of the 21 cm signal, so that the true reach should be actually deeper.

To put this result into some context, we also compare it with a forecast for the sensitivity reach in CMB spectral distortions with {\sf PiXiE}~\cite{Kogut:2011xw} and CMB temperature and polarization anisotropies measurements by a {\sf CORE}-like experiment \cite{Bouchet:2011ck} with sky coverage $f_{\rm sky}$ = 0.70. We find that {\sf PiXiE} would give constraints up to one order of magnitude better than BBN ones, whereas minimal information from an {\em approved} 21 cm experiment would already have comparable sensitivity to the {\em proposed next} generation CMB experiment, making it indeed a very powerful probe. Note that CMB experiment would however still dominate in the energy injection regime corresponding to modifications at the recombination era or shortly thereafter, so the complementarity of different techniques will still be holding in the future.
\section{Conclusion}\label{sec:Concl}
Cosmological probes (BBN, CMB spectral distortions and CMB angular power spectra) lead to stringent bounds on the e.m. decay of exotic particles, spanning about 12 orders of magnitude in abundance and more than 20 orders of magnitude in lifetime. We have re-evaluated these constraints with state-of-the-art tools: Our main results are summarized in Fig.~\ref{Fig:AllConstraints}. 
The main focus of our article has been the computation of CMB anisotropy constraints. For that purpose we have used the most up-to-date available tools, and we expect our bounds to be realistic within the parameter space covered by this analysis. Instead, the BBN and spectral distortions have been reported mostly for comparative purposes. They were derived  under conservative assumptions, and in specific cases the actual bounds may be stronger: for instance one could go beyond the simplistic $\mu$ and $y$ type of  CMB distortions, or take into account the actual photon spectrum leading to photo-hadronic dissociation of light elements, rather than the approximate ``universal spectrum'' limit.

One of the major novelties of our article is the throughout description of the physical effects leading to the CMB bounds, notably for the case of particles whose lifetime is much shorter
than the age of the universe. In particular, CMB is sensitive to particles decaying around the time of recombination, a phenomenon that can lead to peculiar modifications of the power spectra. Based on our results, we also proposed a physical criterion for an approximate ``on-the-spot'' treatment of the problem. This approximation can be used to derive order-of-magnitude estimates of upper bounds, but should be considered with caution if an actual signal was discovered, since it may lead to significant errors with respect to a full calculation of the free electron fraction and CMB power spectra.

The results we obtained can be applied to numerous models, of which we provided a few examples:
We computed the constraints on evaporating low-mass PBHs, whose strength happens to be comparable or stronger than the ones following from extragalactic gamma-ray background limits for masses a couple of orders of magnitude around $10^{15}\,$g. Our results provide for instance an independent cosmological argument excluding that the totality of DM can be made of PBHs lighter than $\sim10^{16.8}\,$g. We have also applied our results to sterile neutrinos: while the past literature has been focusing on masses of a few keV typically invoked for sterile neutrino DM, we have shown that cosmological bounds are also relevant for heavier and more weakly coupled relics, which are usually overlooked.
Finally, we have studied some perspectives for future improvements over the current cosmological constraints. In particular, we sketched why we expect forthcoming 21 cm surveys to have 
some discovery potential, or to improve somewhat over existing bounds. The most promising window to overcome astrophysical uncertainties and aiming at a smoking gun feature is to look for an {\it emission} signal in the ``Cosmic Dawn'' epoch $15\lesssim z\lesssim 30$.
The perspectives to discover this kind of exotic physics are certainly cleaner than those for DM annihilation, see~\cite{Lopez-Honorez:2016sur}. 
We also briefly discussed the good chances for improvements over the current status  achievable with future missions proposed to study CMB spectral distortions ({\sf PiXiE}) and CMB angular power spectrum ({\sf CORE}), as well as their complementarity. 

The above examples do not obviously exhaust the list of applications. Even limiting oneself to the above topics, however, it is clear that each of them would deserve dedicated 
and deeper studies. For instance, given the limitations of the existing study~\cite{Ricotti:2007au}, a revisitation of the cosmological bounds to PBHs of stellar mass (or heavier) is of
high priority. A proper exploration of the cosmologically interesting sterile neutrino parameter space (well beyond the DM-inspired mass-coupling usually considered) remains to be tackled. The actual reach of the 21 cm window should be reassessed with a calculation of the modified power spectrum, a task which we expect will lead to an improved sensitivity to these models. \\
{}\\
{\it Notes added:} While this paper was been finalized, the article~\cite{Slatyer:2016qyl} has appeared, which has some overlap with the topics treated here. Whenever referring to similar models, our results are comparable (see in particular their Fig. 11 and compare with our Fig.~\ref{Fig:AllConstraints}). Also note that our physical prescription for the effective $f(z)$ function, inspired by inspection of the results in Fig.~\ref{Fig:AllConstraints}, finds a technical justification in the principal component analysis developed in~\cite{Slatyer:2016qyl}. For the rest, the two articles are rather complementary in the applications: for instance ref.~\cite{Slatyer:2016qyl} develops a detailed study for very long lived DM, which we chose here to barely treat for the reasons recalled in Sec.~\ref{sec:OtherConstraints}. On the other hand, we also describe applications to non-decaying DM relics, such as primordial black holes, and also cover in better details the constraints on short lifetimes, coming either from CMB or other channels, as well as some perspectives for future cosmological probes.
We also present a throughout description of the physics involved in the CMB effects (Sec.~\ref{sec:CMBconstraints}).

 Also, after our article has appeared in preprint form, ref.~\cite{Clark:2016nst} was finalized. Its content largely overlaps with the one of our Sec.~\ref{PBH}. Apart for minor technical differences and their focus only on the higher range of the PBH masses considered by us, we agree qualitatively with their results. Yet, an appreciable quantitative difference exists, with their bounds being systematically stronger than ours. We believe that the main reason for the difference is that we run our chains by leaving all the cosmological parameters of the extended $\Lambda$CDM model free, while they explicitly state (Sec. IV.C) that they fix the  $\Lambda$CDM parameters to their fiducial values. For a specific mass, they estimate this effect to be a factor 3; we checked that in some cases it can easily reach a factor $\sim 5$. Accounting for this difference, the results are mutually consistent.
\appendix
\acknowledgments

P.D.S. acknowledges support from the Alexander von Humboldt Foundation.
V.P. is supported by the ``Investissements d'avenir, Labex ENIGMASS'',  of  the  French  ANR.
This work has been done thanks to the facilities offered by the Universit\'e Savoie Mont Blanc MUST computing center.
We thank A. Ibarra and J. Chluba for comments on an earlier version of this article. We also thank T. Slatyer and H. Liu for useful discussions.
\bibliographystyle{ieeetr}

\bibliography{biblio}
\end{document}